\documentstyle[preprint,aps,epsf,cite,tighten]{revtex}

\def\beq{\begin{equation}}   
\def\eeq{\end{equation}}

\newcommand{\nc}{\newcommand}
\nc{\grad}{\nabla}
\nc{\tr}{\mathop{\rm tr}}
\nc{\half}{{1\over 2}}
\nc{\third}{{1\over 3}}
\nc{\be}{\begin{equation}}
\nc{\ee}{\end{equation}}
\nc{\bea}{\begin{eqnarray}}
\nc{\eea}{\end{eqnarray}}

\def\Tr{{\rm Tr}}
\nc{\dint}[2]{\int\limits_{#1}^{#2}}
\nc{\D}{\displaystyle}
\nc{\PDT}[1]{\frac{\partial #1}{\partial t}}
\nc{\tw}{\tilde{w}}
\nc{\tg}{\tilde{g}}
\nc{\newcaption}[1]{\centerline{\parbox{5.6in}{\caption{\footnotesize{#1
}}}}}
\def\href#1#2{#2}

\nc{\al}{\alpha}
\nc{\ga}{\gamma}
\nc{\de}{\delta}
\nc{\ep}{\epsilon}
\nc{\ze}{\zeta}
\nc{\et}{\eta}

\nc{\Th}{\Theta}
\nc{\ka}{\kappa}
\nc{\la}{\lambda}
\nc{\rh}{\rho}
\nc{\si}{\sigma}
\nc{\ta}{\tau}
\nc{\up}{\upsilon}
\nc{\ph}{\phi}
\nc{\ch}{\chi}
\nc{\ps}{\psi}
\nc{\om}{\omega}
\nc{\Ga}{\Gamma}
\nc{\De}{\Delta}
\nc{\La}{\Lambda}
\nc{\Si}{\Sigma}
\nc{\Up}{\Upsilon}
\nc{\Ph}{\Phi}
\nc{\Ps}{\Psi}
\nc{\Om}{\Omega}
\nc{\ptl}{\partial}
\nc{\del}{\nabla}
\nc{\ov}{\overline}
\nc{\gsl}{\!\not}
\nc{\bi}[1]{\bibitem{#1}}
\nc{\fr}[2]{\frac{#1}{#2}}
\nc{\gm}{\mbox{$\gamma_{\mu}$}}
\nc{\gn}{\mbox{$\gamma_{\nu}$}}
\nc{\Le}{\mbox{$\fr{1+\gamma_5}{2}$}}
\nc{\Ri}{\mbox{$\fr{1-\gamma_5}{2}$}}
\nc{\GD}{\mbox{$\tilde{G}$}}
\nc{\gf}{\mbox{$\gamma_{5}$}}
\nc{\Ima}{\mbox{Im}}
\nc{\Rea}{\mbox{Re}}

\nc{\av}{\langle \ph\rangle}
\nc{\ntwo}{${\cal N}\!\!=\!2\;$}
\nc{\none}{${\cal N}\!\!=\!1\;$}
\nc{\nfour}{${\cal N}\!\!=\!4\;$}

\newcommand{\bm}[1]{{\mbox{\boldmath $#1$}}}

\def \bi{\bibitem}
\nc{\rf}[1]{(\ref{#1})}
\def \del{\partial}

\begin{document}
\draft
\preprint{
\vbox{\hbox{TPI-MINN-00/08}
      \hbox{UMN-TH-1939/00}
      \hbox{DAMTP-2001-12}
      \hbox{hep-th/0102121}
}}
\title{
Long Range Forces and Supersymmetric Bound States 
}
\author{
Adam Ritz$^{\,a}$\footnote{email: a.ritz@damtp.cam.ac.uk} and
  Arkady Vainshtein$^{\,b}$\footnote{email: vainshtein@mnhep1.hep.umn.edu}
}
\address{
$^a$Department of Applied Mathematics and Theoretical
Physics,
  Centre for Mathematical Sciences, University of Cambridge,
Wilberforce Rd., Cambridge CB3 0WA, UK\\
$^b$Theoretical Physics Institute, University of Minnesota,
116 Church St SE,\\ Minneapolis, MN 55455, USA
}
\maketitle
\thispagestyle{empty}
\setcounter{page}{0}
\begin{abstract}
We consider the long range forces between two BPS particles on the 
Coulomb branch of \ntwo and \nfour supersymmetric gauge
theories. The $1/r$ potential is unambiguously fixed, even at strong
coupling, by the moduli dependence of central charges supported by 
the BPS states.  The effective Coulombic coupling vanishes on
marginal stability curves, while sign changes on crossing these curves
explain the restructuring of the spectrum of composite BPS states.
This restructuring proceeds via the delocalization of the composite
state on approach to the curve of marginal stability.
Therefore the spectrum of BPS  states can be inferred by analyzing 
the submanifolds of the moduli space where
the long range  potential is attractive. 
This method also allows us to find certain 
non-BPS bound states and their stability domains.
As examples, we consider the dissociation of the $W$ boson and 
higher charge dyons at strong coupling in \ntwo SU(2) SYM,
quark-monopole bound states in \ntwo SYM with one flavor,
and composite dyons in \ntwo SU(3) SYM.

\end{abstract}

\newpage

\tableofcontents

\newpage
\section{Introduction}
\label{sec:intro}

Theories with extended supersymmetry admit special shortened massive 
multiplets which preserve some fraction of supersymmetry. In
particular, \ntwo and \nfour supersymmetric gauge theories possess
Bogomol'nyi-Prasad-Sommerfield (BPS) states $|{\bf n}\rangle$ which
along the Coulomb branch are ground states in sectors with a given
set ${\bf n}=\{{\rm n}_i\}$ of conserved  U(1) charges. As shown in 
\cite{wo,ho}, BPS states
preserve some fraction of supersymmetry because they support at least one
central charge ${\cal Z}$ 
of the superalgebra, ${\cal Z}|{\bf n}\rangle={\cal
Z}_{\bf n}|{\bf n}\rangle$. Moreover, BPS states inherit special 
stability properties from the constraint that their masses  are fixed 
in terms of their central charges $M_{\bf n}=|{\cal Z}_{\bf n}|$.  
Such a state is
thus stable with respect to decay into two particles, $|{\bf
n}_1\rangle$ and  $|{\bf n}_2\rangle$, with ${\bf n}={\bf n}_1+{\bf n}_2$.
Indeed, since the central charges are linear in $\{{\rm n}_i\}$,
$M_{\bf n}=|{\cal Z}_{{\bf n}_1}+{\cal Z}_{{\bf n}_2}|\le M_{{\bf n}_1}
+M_{{\bf n}_2} $. However, the central charges generically depend on the 
moduli (and/or parameters) in the theory, and so 
there can be special submanifolds where a BPS state is only marginally
stable, i.e. $M_{\bf n}=M_{{\bf n}_1}+M_{{\bf n}_2}$.
These {\it curves of marginal stability} (CMS) in the moduli (and/or parameter)
space are then of interest as they are the only regimes where a
discontinuous change in the spectrum of BPS states is possible.

The existence of marginal stability
submanifolds is quite generic in theories where the
superalgebra can support central charges. They have been observed
in two dimensional models \cite{Vafa}, in the particle spectrum
of \ntwo and \nfour SYM \cite{sw,n2cms,fb1,bf2,bs,bf3}, and in related 
Type IIB string junctions \cite{bergman,Wsj}. More generally, they
arise in the consideration of the D-brane spectrum in string
compactifications on manifolds with nontrivial cycles \cite{cyau}.
CMS curves may also arise for extended objects such as BPS
domain walls \cite{SmVe,Sm} in theories without extended
supersymmetry.

While the existence and position of CMS curves is straightforwardly deduced
from the BPS mass formula and the moduli dependence of the central charges
${\cal Z}_{\bf n}\,$, it is not immediately apparent whether 
a discontinuity in the spectrum does occur, and if so on which side 
of the CMS do the BPS states actually exist, or indeed whether they exist
at all. However, Bilal and Ferrari \cite{fb1,bf2,bf3} were able to
show that, for \ntwo  theories with gauge group SU(2), a full reconstruction 
of the BPS spectrum  accounting for its discontinuities on the CMS can 
in fact be deduced from the moduli dependence of the central charges. 
The derivation is not dynamical, it is based on 
continuity, symmetries, and knowledge of the singularities present in
the Seiberg-Witten \cite{sw} low energy solution.  A somewhat
more dynamical construction of the spectrum
was subsequently made via an examination of Type IIB 
string junctions \cite{bergman,Wsj} in appropriate backgrounds.

In this paper we will show that the dynamics which governs the
formation of composite BPS states, and describes the corresponding
discontinuities of the spectrum, is quite straightforward. In particular,
there are long range Coulombic forces between BPS
particles due to massless electrostatic, magnetostatic, and 
scalar exchanges, and it is the sign of this long range potential
which determines the presence or otherwise of bound states. Moreover,
this potential is unambiguously fixed by the moduli 
dependence of the central charges and proves to be calculable in all
regions of the moduli space.

The importance of this potential relies on the central idea of our
work together with Shifman and Voloshin \cite{rsvv}, which is that
in the near CMS region ``composite'' BPS particles can be viewed as
weakly bound states of ``primary'' BPS particles, and moreover that these 
composite states {\it delocalize} on approach to the CMS. 
In 3+1D this delocalization follows from the structure of the BPS mass
formula and the constraint that the  Coulombic 
attraction must vanish on the CMS. Generically the effective 
Coulombic coupling changes sign upon crossing the CMS, and thus
attraction changes to repulsion and the ``composite'' BPS state 
is removed from the physical spectrum. The work presented here 
generalizes the analysis of \cite{rsvv} to strong coupling.

Delocalization near the CMS allows us to treat the BPS constituents
as point-like charges interacting via long range forces and 
described by supersymmetric quantum mechanics (SQM) \cite{EWi}. 
This approximation is valid when the binding energy
$E_{\rm bind}=M_{{\bf n}_1}+M_{{\bf n}_2}-M_{{\bf n}}$ is much less
then the constituent masses $M_{{\bf n}_1}\,$, $M_{{\bf n}_2}\,$. 
Crucially, this constraint can be satisfied for any CMS
curve, including those at strong coupling, provided we keep away from the
singularities where the BPS constituents become massless, $M_{{\bf n}_1, {\bf
n}_2}=0$. Therefore, by evaluating the long range potential and observing in 
which regions of the moduli space it is attractive or repulsive, we
can deduce the spectrum of BPS and also non-BPS bound states composed
from BPS constituents. While it might seem 
rather remarkable that such an approximation makes sense in the 
strong coupling region, the crucial point is that since the 
constituent states delocalize on approach to the CMS, the interaction
between constituents 
can be made arbitrarily weak, and thus the approximation arbitrarily
good by moving closer to the CMS itself. 

The physical picture is that as a composite state is moved closer to
its CMS curve it behaves like a weakly bound Coulomb system, and its 
constituents dissociate on the CMS itself. Notice that 
in terms of the scattering amplitude for the $|{\bf n}_1\rangle$ and  
$|{\bf n}_2\rangle$ particles one still observes 
continuity on crossing the CMS in the following sense. In the complex energy
plane the pole singularity associated with the bound state  
moves to the beginning of the continuum cut when the moduli approach 
the CMS. After crossing the CMS the singularity goes back along the
same line but on the second sheet of the Riemann surface.  A 
singularity of this kind, close to the continuum and on 
the second sheet is known as
a virtual state (or virtual level). This situation is quite distinct 
from the formation of a resonance, and for this
reason we prefer to call the phenomenon on crossing the CMS
a ``dissociation'' (or delocalization) into primary particles 
rather then a ``decay'', the term often used in the literature. 

The discussion above was framed in terms of \ntwo supersymmetry with 
one complex central charge ${\cal Z}$. However, by considering
a massless \ntwo  hypermultiplet in the adjoint
representation we can include the \nfour case. It is well known that 
duality \cite{mo} predicts a BPS spectrum for \nfour SYM which 
is considerably richer than that of 
\ntwo SYM, e.g. for gauge group SU(2) there are
BPS states with magnetic charge two (and nonvanishing
co-prime electric charge) present in \nfour SYM, which are not in
the \ntwo BPS spectrum. Making use of the Coulombic potential, we can 
understand this behavior as follows: breaking \nfour to \ntwo
actually preserves bound states with magnetic charge two, in the 
semiclassical domain of the moduli space, but these states are 
{\em not} BPS states of \ntwo$\!$. Thus, the total BPS plus non-BPS
spectrum behaves continuously under breaking to \ntwo$\!$. Moreover, 
we find that these non-BPS states possess a new marginal stability 
curve that we call the \mbox{``non-BPS CMS''}, or ``nCMS'', 
and are not present inside a certain strong coupling submanifold
of the moduli space bounded in part by this curve. 
This result about the spectrum of certain non-BPS
states of higher magnetic charge in \ntwo SU(2) 
SYM is in agreement with the arguments of Bergman \cite{B2} made using
the string junction construction. 

The layout of the paper is as follows. In Sec.~\ref{sec:2} we describe the 
calculation of the long range potential between two BPS states due to
massless exchange. We then apply this potential to a number of examples
illustrating how it can be used to deduce features of the BPS and
non-BPS bound state spectrum in \nfour and \ntwo SYM. In Sec.~\ref{sec:3}, we
consider pure \nfour and \ntwo SYM with gauge group SU(2), 
reproducing the expected BPS
spectrum at weak and strong coupling, 
and deducing the presence of many non-BPS states in \ntwo
SYM. In Sec.~\ref{sec:4}, we consider \ntwo SYM with gauge group SU(2) and one
massive flavor, which provides a convenient example for considering
the effect of the Argyres-Douglas point \cite{ad} on the behavior
of the potential in the strong coupling region. Quark-monopole
bound states were considered in similar theories in \cite{bf3}, and 
at weak coupling in \cite{henn}.
In Sec.~\ref{sec:5}, we generalize the construction to gauge group 
SU(3). In this system CMS curves extend to the weak coupling region, 
and we study the restructuring of the spectrum of dyons in the
semiclassical regime on breaking \nfour to \ntwo$\!\!$.
Composite dyons in this theory have recently been considered in
the moduli space approximation in 
\cite{ly,bhlms,tong,blly,bly,bl,bly2,n2quant,rsvv,gkly}. 
Finally, in Sec.~\ref{sec:6}, we finish 
with some concluding remarks and directions for future work.

While this work was being written up we received an
interesting preprint \cite{argyres} by Argyres and Narayan 
which also discusses the discontinuities of the BPS spectrum 
in \ntwo and \mbox{\nfour} SYM. Our work shares with \cite{argyres}
the general picture \cite{rsvv} of loosely bound 
BPS states near the CMS. However, the approach of Argyres and 
Narayan is technically different from ours, as they describe 
equilibrium configurations rather than the interactions, and their
construction makes a bridge to the string web picture (see also
\cite{kk1}).  
It would be interesting to explore the possible connections between these 
approaches in more detail.

\section{The BPS spectrum and long range potentials}
\label{sec:2}

In this section we discuss in detail the form of the long range
potential between BPS sources, and obtain a convenient representation
near the CMS. 

\subsection{Generic Constraints}

We begin by recalling some of the general constraints imposed on 
long range interactions between BPS states in SYM, as discussed
in \cite{rsvv}. First, as mentioned above, it is clear that if we
consider the dynamics of two BPS particles with masses $M_{{\bf n}_1}$ and
$M_{{\bf n}_2}$ sufficiently near the CMS for the composite state $|{\bf
n}\rangle=|{\bf n}_1+{\bf n}_2\rangle$ with mass
$M_{\bf n}\,$, then the  binding energy $E_{\rm bind}$ 
(and, consequently, the kinetic energy)  can be made much  smaller than the
masses $M_{{\bf n}_1}$ and $M_{{\bf n}_2}$ of the primary states 
themselves. i.e. by moving near the CMS we
have,
\be
 \ep = \frac{E_{\rm bind}}{M_r} \ll 1\,,\qquad
E_{\rm bind}=M_{{\bf n}_1}+M_{{\bf n}_2}-M_{{\bf n}}\,,\qquad
M_r=\frac{M_{{\bf n}_1} M_{{\bf n}_2}}{M_{{\bf n}_1} + M_{{\bf
n}_2}}\,,
 \label{const}
\ee
where $M_r$ is the reduced mass of the two particles. This is a 
restriction that we need to keep away from points in moduli space
where the BPS states are massless. However, this is not a significant
constraint as such submanifolds are of higher co-dimension than the
CMS curves. Therefore we can always consider a regime near the
CMS where the dynamics
describing the composite state is manifestly nonrelativistic, i.e.  where
$\ep \ll 1$. In this regime we can legitimately ignore the full 
microscopic theory and study the effective (supersymmetric)
quantum mechanics associated with the collective coordinates of the
BPS states. It is important that this argument relies only on
having $\ep \ll 1$ and is therefore quite general; it applies whether 
or not the underlying theory is strongly or weakly coupled.

Moreover, in 3+1D this point of view can be pushed one stage further
since the notion of marginal stability addressed here can 
be rephrased at the quantum mechanical level
as a question about the presence or otherwise of bound states.
The form of the BPS mass formula indicates that as we move
near a CMS curve, bound states are formed via an attractive
coupling that can be made arbitrarily small.  
In contrast with 1+1 and 2+1 dimensions 
where any arbitrarily small attraction is sufficient to form a bound 
state, in 3+1D only long range potentials can form a 
bound state for arbitrarily small coupling, 
and thus we deduce that such forces must be present.  In other
words, bound state formation near the CMS depends only on long
range forces, and is therefore insensitive to detailed issues 
about short range interactions between the core structures of 
the BPS states. Consequently, we can simplify the effective
description by treating the primary states as point-like
particles interacting only via long range forces. This is clearly 
an abstraction but these arguments indicate that it is quite sufficient 
for answering questions about the presence of bound states.

The consistency of this argument can be verified by noting
that Coulombic systems associated with an attractive $1/r$-type
potential, 
\be
 V(r) = - \frac{f}{4\pi r}\,,
\label{Vc}
\ee
in the limit where the coupling $f$ becomes small
possess towers of closely spaced  bound
states, only the lowest of which can be BPS saturated. In contrast, we know
from the BPS mass formula that on the CMS the lowest level in the
tower must reach the continuum. This can only happen if the 
effective coupling $f$, which is a function of the 
moduli, vanishes on the CMS. 

A corollary of these constraints is that the composite BPS
configuration, viewed as a bound state of primary constituents,
must delocalize on approach to the CMS.  An example of such delocalization 
at the classical level of static equilibrium configurations is
provided by the semiclassical system of two dyons with charges under
different U(1) subgroups of the Cartan subalgebra in \ntwo SYM with
gauge group SU(3) \cite{ly,blly,bly,n2quant,gkly}, as discussed from this
standpoint in \cite{rsvv}.  In this system, besides the Coulombic 
potential (\ref{Vc}) with a coupling $f$ which vanishes on the CMS,
there is also a subleading repulsive $1/r^2$ potential which 
does not vanish on the CMS. This leads to a diverging
equilibrium separation on approach to the CMS, which becomes
infinite on the CMS itself.

With this information in hand, it follows that the discontinuity
in the spectrum on crossing a CMS can be inferred purely from the 
leading Coulombic potential between constituent states. The rest of
this section is devoted to constructing this potential for a generic
\ntwo theory (we include \nfour SYM by allowing for the presence of 
a massless adjoint hypermultiplet).

\subsection{Evaluating the Long Range Potential}

Recall first of all that the mass of a BPS state is given by the 
absolute value of the corresponding central charge ${\cal Z}$
\be
 M_{\bf n} =\left|{\cal Z}_{\bf n}\right|\,, \qquad  
{\cal Z}\,|{\bf n}\rangle={\cal Z}_{\bf n}\,|{\bf n}\rangle\,.
\label{Mbps}
\ee
The central charge ${\cal Z}$ arising in the \ntwo supersymmetry 
algebra is a linear combination of conserved U(1) charges $\{Q_i\}\,$ 
with coefficients $c^i$ which are functions of the vacuum moduli $u_k$,
\begin{equation}
  \label{centc}
  {\cal Z}=\sqrt{2}\, \sum_i Q_i\, c^i(u)\,.
\end{equation}
We label the state $|{\bf n}\rangle$ with a set of
parameters ${\bf n}=\{{\rm n}_i\}$ which are the charges of the state
with respect to the  $\{Q_i\}\,$,
\begin{equation}
  \label{charges}
Q_i\, |{\bf n}\rangle = {\rm n}_i\, |{\bf n}\rangle\,.
\end{equation}
Thus,
\begin{equation}
  \label{centc1}
{\cal Z}_{\bf n}=\sqrt{2}\,\sum_i {\rm n}_i\, c^i(u)\,.
\end{equation}

For \ntwo theories with gauge group SU($N$) maximally broken to 
U(1)$^{N-1}$, the charges ${\bf n}$ we deal with can be represented in 
the following form ${\bf n}=\{n_{E}^{a},n_{Ma},s^f\}$, where 
the index $a=1,\ldots,N-1$ refers to a suitable basis in 
the Cartan subalgebra. 
In this set, $n_{E}^{a}$ and $n_{Ma}$ refer 
respectively to the lattice of electric charges 
and the dual lattice of magnetic charges.  
In our conventions, $n_{E}^{a}$  and $n_{Ma}$ are integers or
half-integers\,\footnote{In this respect our conventions differ from 
Seiberg and Witten \cite{sw}, who modified the definition of the 
corresponding vev to make $n_{E}^{a}$, for example, integral.  
Another difference in conventions is the $\sqrt{2}$ in the definition of the 
central charge ${\cal Z}$.}. 
Note that half-integer electric charges arise  in the
presence of fundamental matter, while half-integer magnetic
charges appear in a related way, as discussed below, for higher rank gauge 
groups. Finally, $s^f$ are integral flavor charges 
associated with U(1) symmetries of the additional  hypermultiplets.
The central charge may then be written in the form
\be
{\cal Z}_{\{n_{E}^{a},n_{Ma},s^f\}} = \sqrt{2}
\left[n_{E}^{a}\, a_a(u) + n_{Ma}\, a_{D}^{a}(u)\right] + s^f m_f\,, 
\label{BPSmass}
\ee
where $a_a$ and $a_{D}^{a}$ are functions of the vacuum 
moduli $u_k$ which are defined via powers of the adjoint 
scalar field $\Phi$ as $u_k=\langle \Tr\,\Ph^k\rangle+$products of 
lower order Casimirs, and $m_f$ is the mass of the $f$-th hypermultiplet.

The low-energy effective Lagrangian for the \ntwo SYM massless fields, 
written in terms of \none superfields, has the following 
form \cite{SUN},
\begin{equation}
{\cal L}_{\rm eff}=
\frac{1}{8\pi}\,{\rm Im}\!\int\!\!  {\rm d}^2\theta \,W_a W_D^a+
\frac{1}{4\pi}\,{\rm Im}\!\int\!\!  {\rm d}^2\theta {\rm d}^2\bar\theta
\,\bar A_a A_{D}^{a}
\label{leff}
\end{equation}
where $\langle A_a \rangle=a_a(u)$, 
$\langle A_{D}^{a} \rangle=a_{D}^{a}(u)$, and $W_D$ and 
$A_{D}^{a}(A)$ are defined 
by the prepotential ${\cal F}(A)$, 
\be
W_D^a= \tau^{ab}(A)\,W_b\,,\qquad
A_{D}^{a}(A)=\frac{\partial{\cal  F}(A)}{\partial A_a}\,, \qquad 
\tau^{ab}(A)=\frac{\partial^2{\cal F}(A)}{\partial A_a \partial A_b}\;.
\label{defAW}
\ee

Thus far we have not specified a basis in the Cartan subalgebra.
However, the integrality properties of the charges discussed above 
are assured by choosing a basis in which the index 
$a=1,\ldots,N-1$ effectively enumerates the simple roots $\{\bm{\beta}_a\}$
(for the electric charges) and the fundamental weights
$\{\bm{\om}^a\}$ (for the magnetic charges) of 
the algebra (see e.g. \cite{klt}). This
structure is natural recalling that the simple roots and fundamental
weights are orthonormal\,\footnote{We will not distinguish roots and
co-roots for SU($N$).}, $\bm{\om}^{a}\cdot\bm{\beta}_b=\de^a_b$,
and generate dual lattices. They are related via the Cartan matrix,
which for SU($N$) takes the form 
$C_{ab}=\bbox{\beta}_{a}\cdot\bbox{\beta}_{b}$ and in our
normalization $C_{aa}=1$ and $C_{aa\pm1}=-1/2$. The transition
between, say, the electric charge lattice and its dual is accomplished via 
$\bbox{\beta}_{a}=C_{ab}\,\bm{\om}^{b}$. 

This index convention is well adapted to the duality structure of the
low energy effective theory. In the semiclassical regime the
coupling matrix $\ta^{ab}$ defined in Eq.~(\ref{defAW}) is proportional 
to the Cartan matrix, 
\begin{equation}
\tau^{ab}\to \,\tau_0\,C^{ab}\,,\qquad  \tau_0=\frac{4\pi
i}{g_0^2}+\frac{\theta_0}{2\pi}\,,
\end{equation}
and we can write 
$a_a=\langle\bm{\Phi}\cdot\bm{\beta}_{a}\rangle$ and 
$a_{D}^{a}=\ta^{ab}a_b$. It is then helpful to define magnetic charges
with a lifted index $n_M^a=C^{ab}n_{Mb}$ which are referred to the
basis of simple roots \cite{gno,weinberg}. This leads to the conventional 
semiclassical form for the central charge, 
\be
{\cal Z}_{\{n^{a}_{E},n_{M}^a,s^f\}}^{\rm cl} = 
\sqrt{2}\,[n^{a}_{E} +\tau_0\, n_{M}^a]\, a_a +s^f\,m_f\,.
 \label{Zsc}
\ee
Note that while the magnetic charges $n_M^a$ are integral, they 
lead to charges $n_{Ma}=C_{ab}n_M^a$ on the dual lattice which contain
half-integral components.

The sources for the massless fields described by (\ref{leff})
are massive hypermultiplets, each described by a pair 
$X$, $\tilde X$ of \none chiral superfields. The 
low energy effective  Lagrangian for the hypermultiplets, which
includes their interaction with the massless fields, has
the form
\begin{equation}
{\cal L}_{\rm hyp}=
 \sum_X\left\{\!\int\!\!{\rm d}^2\theta{\rm d}^2\bar\theta\left[\bar X{\rm
    e}^{n_{E}^{a} V_a} {\rm e}^{n_{Ma} V^a_D}X
+\bar{\tilde X}{\rm e}^{-n_{E}^{a} V_a }{\rm e}^{-n_{Ma} V^a_D}\tilde X\right]
 + 2\,{\rm Re}\!\int\!\!{\rm d}^2\theta\,\tilde{X} {\cal Z}_X(A)
 X\right\},
\label{hypeff}
\end{equation}
where $n_{E}^{a}$, $n_{Ma}$ are the electric and 
magnetic charges of the field $X$, the superfield $V_D$ is dual 
to $V$, and ${\cal Z}_X(A)$ is given by
Eq.~(\ref{BPSmass}) in which ${\cal Z}$ is viewed as a function of $A_a$. 

\begin{figure}[ht]
\epsfxsize=5cm
\centerline{%
   \epsfbox{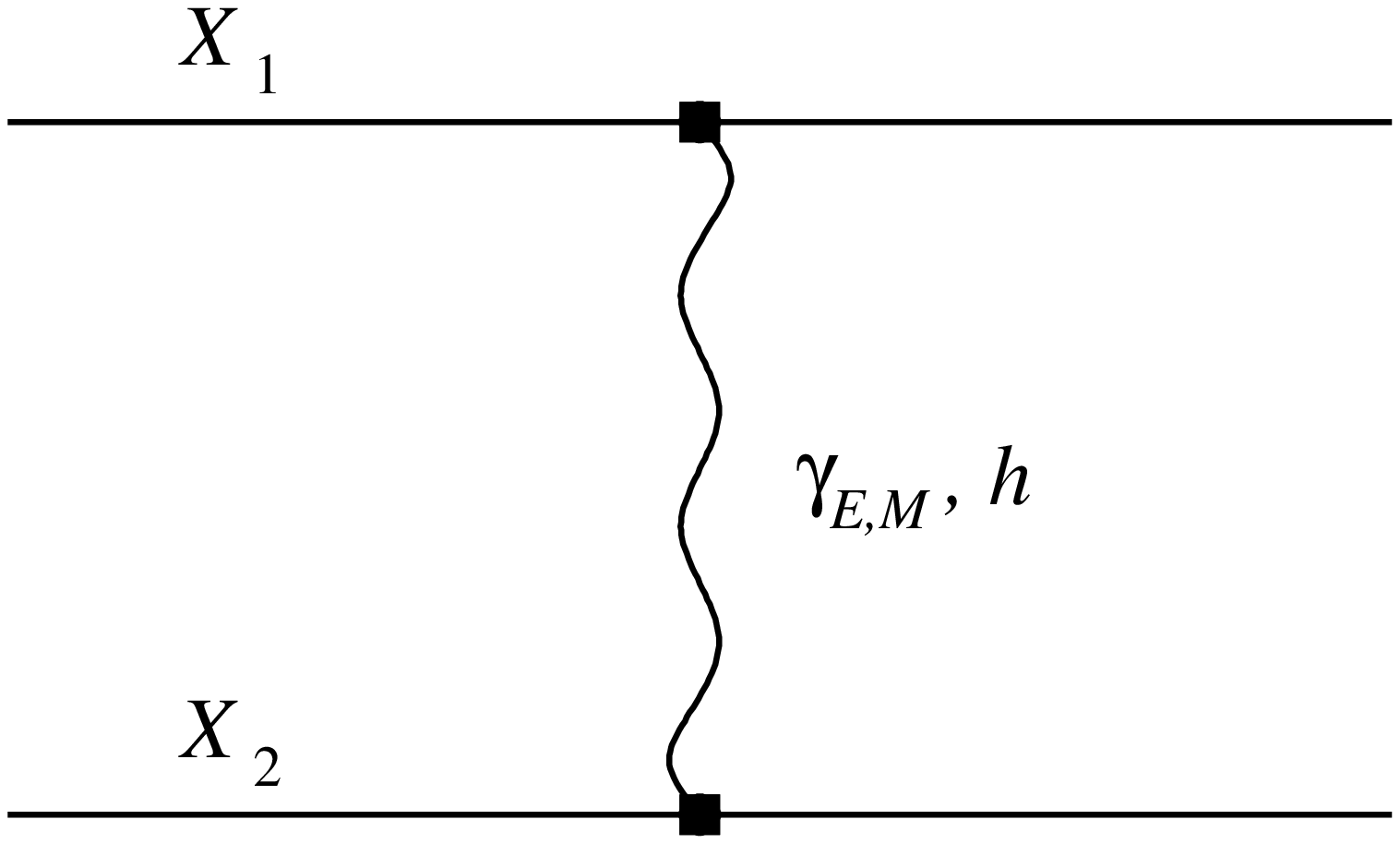}%
         }
 \vspace*{0.2in}
 \newcaption{The tree level massless exchanges 
leading to the long range potential between the BPS 
states $X_1$ and $X_2$. $\ga_{E,M}$ refers to the electrostatic and
magnetostatic terms, while $h$ denotes the scalar field quanta.
\label{fig:vertex}}
\end{figure}

The Lagrangians (\ref{leff}) and (\ref{hypeff}) together with the expression 
(\ref{BPSmass}) for the central charge contain everything required 
to determine the long range interaction between two static 
BPS particles due to the massless exchanges shown schematically 
in Fig.~\ref{fig:vertex}. Indeed, ${\cal L}_{\rm
hyp}$ describes the couplings between the hypermultiplets and the
massless fields while ${\cal L}_{\rm eff}$ defines the massless 
propagators. Note that the point-particle approximation allows
us to avoid any subtleties with considering states which are not
mutually local within a field theoretic description. Strictly, we 
are calculating the classical energy for a 
system of heavy particles with electric and magnetic charges. 
\firstfigfalse
The $1/r$ potential is obtained as the sum of three terms,
\be
 V(r) = V_{E}(r) + V_{M}(r) + V_S(r), \label{Vr}
\ee
referring to electrostatic $V_{E}$, magnetostatic $V_{M}$, 
and scalar $V_S$ exchange, in correspondence with the 
contributions to the tree level diagram in
Fig.~\ref{fig:vertex} where $\gamma_a$ and $h_a$ are 
the intermediate massless particles. 
Let us start with the electrostatic interaction. 
When all magnetic charges $n_{Ma}$
are zero we  can straightforwardly read off its 
form from the low energy effective
Lagrangians (\ref{leff}), (\ref{hypeff}). The part of 
the Lagrangian (\ref{leff}) describing the electric field is
\begin{equation}
{\cal L}_E=-\frac{1}{2}\, g^{ab}\,\vec E_a\,\vec E_b\,,
\qquad g^{ab}=\frac{{\rm Im}\,\tau^{ab}}{4\pi}\,.
\label{esta}
\end{equation}
The electrostatic interaction $V_E$ is then 
\begin{equation}
  \label{estat}
  V_E=\frac{1}{4\pi r}\; n^{(1)a}_{E}\,g_{ab}\;n^{(2)b}_{E}\,, \qquad
g_{ab}=\left[\frac{4\pi}{{\rm Im}\,\tau}\right]_{ab} \,,
\end{equation}
where $g_{ab}$ is the inverse of the metric  $g^{ab}$ defined in
Eq.~(\ref{esta}). 

From the dual description it is clear that when only magnetic
charges are present the interaction is 
\begin{equation}
  \label{mstat}
  V_M=\frac{1}{4\pi r}\;n^{(1)}_{Ma}\,g^{ab}\;
n^{(2)}_{Mb}\,.
\end{equation}
When both electric and magnetic charges are present, besides summing up
the potentials (\ref{estat}) and (\ref{mstat}), one also needs to account for 
the Witten effect~\cite{weffect}, i.e. the electric
charges $n_{Ea}$ in Eq.~(\ref{estat}) should be substituted 
by  $n_{Ea}+n_M^b\,{\rm Re}\,\tau_{ab}$.  In terms of massless 
exchange, the Witten effect implies a mixing
between the fields $V_a$ and $V_D^a\,$. The mixing arises because
the field $V_D^a$ contains an electric part ${\rm Re}\,\tau^{ab}\,V_b$
along with the magnetic part which does not mix with $V_a$.
Thus, for the sum  of $V_E$ and $V_M$ we get
\begin{equation}
  \label{emstat}
  V_E+V_M=\frac{1}{4\pi r}\; {\rm
    Re}\left\{\left(n^{(1)a}_{E}+n^{(1)}_{Mc}\,\tau^{ca}\right)g_{ab}
 \left(n^{(2)b}_{E}+\bar\tau^{bd}\, n^{(2)}_{Md}\right)\right\}\,,
\end{equation}
where the bar denotes complex conjugation.
Noting that $\partial{\cal Z}/\partial a_a\!=\!\!\sqrt{2}\,(n^{a}_{E}
+n_{Mc}\tau^{ca})$
we can rewrite this in the form
\begin{equation}
V_E+V_M
=\frac{1}{8\pi r}\,{\rm Re}\left\{g_{ab}\,\frac{\partial 
{\cal Z}_1}{\partial a_a} 
\frac{\partial\bar{\cal Z}_2}{\partial\bar a_b}
  \right\}\,.
\label{emstat1}
\end{equation}
The ``effective charge'' in this expression can be thought of as the
dispersion between the two  central charges ${\cal Z}_1$ and ${\cal Z}_2$ as
measured via the  K\"ahler metric.

We now turn to the scalar exchange potential which will
be the object of interest for the remainder of this subsection.
The non-supersymmetric scalar potential was first discussed 
from this point of view by Montonen and Olive \cite{mo} where it was 
used to verify the absence of static forces between two identical $W$ bosons. 
It depends on the Yukawa couplings associated with interaction of the massless 
fields  $A^a$ with the BPS states $X_i$.
Importantly, for BPS particles these couplings 
are fixed by supersymmetry and can be deduced by expansion 
of the corresponding central charges ${\cal Z}_X$. Indeed,
these couplings follow from the superpotential term 
in ${\cal L}_{\rm hyp}$ (the last term in Eq.~(\ref{hypeff})). 
When $A_a$ is substituted by its vacuum
value $a_a(u)$ this term gives the mass for the field $X$. Introducing 
the deviations $H_a=A_a-a_a(u)$ from the vevs and 
expanding the central charge ${\cal Z}_X$,
\be
 {\cal Z}_X(A) = {\cal Z}_X(a) 
       +\frac{\partial {\cal Z}_X(a)}{\partial a_a}\,H_a
+ \cdots\,, 
\ee
we deduce the couplings of interest,
\begin{eqnarray}
&& 2\,{\rm Re}\left\{
\frac{\partial {\cal Z}_X(a)}{\partial a_a}
\!\int\!\!{\rm d}^2\theta\,
H_a \tilde X\, X\right\} \nonumber\\[1mm]
&&= -2\,{\rm Re}\left\{\bar {\cal Z}_X(a)
\frac{\partial {\cal Z}_X(a)}{\partial a_a}\,h_a\right\} 
\left(|x|^2+|\tilde x|^2\right)
+\mbox{fermionic terms}\,,
\end{eqnarray}
where $h_a$, $x$ and $\tilde x$ are the 
bosonic components of $H_a$,
$X$ and $\tilde X$.

We also require the propagator for 
the deviations $h_a$. Expanding the second term in Eq.~(\ref{leff}) we 
find, in the quadratic approximation,
the Lagrangian for the $h_a$ fields,
\be
\label{metric}
{\cal L}_h= g^{ab}\partial_\mu \bar h_a\, \partial^\mu 
h_b\,,
\ee
where the metric $g^{ab}$ is defined in Eq.~(\ref{esta}).
The propagator for the scalars $h_a$ is proportional to the inverse
metric $g_{ab}$,
\be
\int\! {\rm d}^4 x\, {\rm e}^{ikx} \left\langle\, T\{ h_a(x),
\bar h_b(0)\}\right\rangle= i\,\frac{g_{ab}}{k^2}\,.
\ee

With this information in hand, we can straightforwardly write down the
static long range potential between the two BPS sources corresponding
to the tree level scalar exchange in Fig.~1,
\be
\label{scalp}
 V_S(r) = -\frac{1}{8\pi r}\,{\rm Re}\left\{\left(g_{ab}\,\frac{\partial 
{\cal Z}_1}{\partial a_a} 
\frac{\partial\bar{\cal Z}_2}{\partial\bar a_b}
\right)
 \left(\frac{\bar{{\cal Z}}_1{\cal Z}_2}{|{\cal Z}_1|\,|{\cal
Z}_2|}\right)
  \right\}.
\ee
We see that the scalar exchange differs from 
$V_E +V_M$, given in Eq.~(\ref{emstat1}), only by presence of
a phase factor naturally  interpreted as the relative 
orientation of the two central charges.  
Furthermore, we observe that, accounting for the presence 
of the metric $g_{ab}$, this expression is reparametrization 
invariant with respect to the choice of the
exchanged scalar field. In other words, our choice of that scalar as
$a_a$ was purely for convenience and the result would be the same
if we had chosen any other scalar, e.g. 
deviations of moduli $u_k$, or a duality related combination such 
as $a^a_D$, which is an important consistency check. 
The use of  $a_a$  is nonetheless convenient because this is the field
which lies in the same supermultiplet as the photon. 

For the total Coulombic potential $V=V_E+V_M+V_S$, we find
\begin{eqnarray}
\label{scalp1}
 V(r) &= &\frac{1}{8\pi r}\,{\rm Re}\left\{\left(g_{ab}\,\frac{\partial 
{\cal Z}_1}{\partial a_a} 
\frac{\partial\bar{\cal Z}_2}{\partial\bar a_b}
\right)
 \left(1-\frac{\bar{{\cal Z}}_1{\cal Z}_2}{|{\cal Z}_1|\,|{\cal
Z}_2|}\right) \right\}\nonumber\\[1mm]
&= &
\frac{1}{r}\;{\rm Re}\left\{\left(n^{(1)a}_{E}+
n^{(1)}_{Mc}\tau^{ca}\right)
\left[\frac{1}{{\rm Im}\,\tau}\right]_{ab}\,
 \left(n^{(2)b}_{E}+\bar\tau^{bd}\, n^{(2)}_{Md}\right)
 \left(1-\frac{\bar{{\cal Z}}_1{\cal Z}_2}{|{\cal Z}_1|\,|{\cal
Z}_2|}\right)
  \right\}.
\end{eqnarray}
The first term in the final parentheses is due to $V_E+V_M$ while the
relative phase between the central charges comes from $V_S$.
The potential (\ref{scalp1}) constitutes the long
range interaction between any two BPS particles. The  generic
feature which is immediately seen from Eq.~(\ref{scalp1}) is that the
potential vanishes on the CMS, where ${\cal Z}_1$ and ${\cal Z}_2$
are aligned, i.e. where ${\rm Im}(\bar {\cal Z}_1 {\cal Z}_2)=0$ 
and ${\rm Re}(\bar{\cal Z}_1 {\cal Z}_2)>0$.  
Let us introduce the angle of disalignment $\omega$ between the
two central charges,
\begin{equation}
{\rm e}^{i\omega}= 
\frac{\bar{{\cal Z}}_1{\cal Z}_2}{|{\cal Z}_1|\,|{\cal Z}_2|}\,.
\label{omd}
\end{equation}
Then $V(r)$ can be rewritten as
\begin{eqnarray}
V(r)&=& \frac{1-\cos\omega}{r}\left\{
\left(n^{(1)a}_{E}+
n^{(1)}_{Mc}\,{\rm Re}\,\tau^{ca}\right)
\left[\frac{1}{{\rm Im}\,\tau}\right]_{ab}
 \left(n^{(2)b}_{E}+{\rm Re}\,\tau^{bd}\, n^{(2)}_{Md}\right)+
n^{(1)}_{Ma}\,{\rm Im}\,\tau^{ab}\,n^{(2)}_{Mb}\,
\right\}\nonumber\\[1mm]
&&-\,\frac{\sin\omega}{r}\left(n^{(1)a}_{E} n^{(2)}_{Ma}-n^{(1)}_{Ma}
    n^{(2)a}_{E} \right)\,,
\label{scalp2}
\end{eqnarray}
which provides a convenient form for consideration of the
near-CMS regime.

\subsection{The Potential in the Near-CMS Region}

In the vicinity of the CMS, i.e. for $|\omega|\ll 1$, 
the last term in Eq.~(\ref{scalp2}) is dominant and we arrive
at a remarkably simple expression for the 
Coulombic potential in the near-CMS region,
\begin{eqnarray}
V_{\rm CMS}&=&-\frac{\omega}{r} \left(n^{(1)a}_{E} n^{(2)}_{Ma}-n^{(1)}_{Ma}
    n^{(2)a}_{E} \right) +{\cal O}(\omega^2)\nonumber\\[1mm]
&=&-\frac{1}{r}\, {\rm Im} \frac{\bar{{\cal Z}}_1{\cal Z}_2}{|{\cal
    Z}_1| \,|{\cal Z}_2|}\left(n^{(1)a}_{E} n^{(2)}_{Ma}-n^{(1)}_{Ma}
    n^{(2)a}_{E} \right) +{\cal O}(\omega^2)\,.
 \label{Vcms}
\end{eqnarray}
This  potential contains no explicit dependence on the 
couplings $\tau^{ab}$, and thus on the K\"ahler metric, 
the dependence on moduli enters only via the
angle of disalignment $\omega$. The potential vanishes on the CMS where
$\om=0$, while for existence of the 
composite ${\bf n}^{(1)}+{\bf n}^{(2)}$ BPS state the potential
$V_{\rm CMS}$ should be attractive. It is clear that this will be
the case on only one side of the CMS, and
thus on the other side the constituents feel a long range 
repulsion and the BPS state ceases to exist.

At this point we should mention that the expression (\ref{Vcms}) is
the near CMS form for a generic charge sector, where the
product $(n^{(1)}_{Ea} n^{(2)a}_M-n^{(1)a}_M n^{(2)}_{Ea})$ is
nonzero. In  sectors for which this combination vanishes 
(e.g. if the constituents are purely electrically or purely magnetically
charged) the total potential generically starts at 
${\cal O}(\om^2)$ as is apparent from Eq.~(\ref{scalp2}). 

In the following sections we will consider some illustrative 
examples which exhibit the basic phenomena of interest.

\section{Spectrum in SU(2) Yang-Mills theory}
\label{sec:3}

In this section we use the long range potential to determine the spectrum of 
bound states for an SU(2) gauge group. Let us consider \ntwo SU(2) theory
with one adjoint hypermultiplet. The expression for the central charge
in this case has the form
\begin{equation}
{\cal Z}_{\{n_E,n_M,s\}}=\sqrt{2}
\left[\,n_{E}\, a(u) + n_{M}\, a_{D}(u)\right] + s\, m_h\,, 
\label{zs2}
\end{equation}
where $m_h$ is the mass of the hypermultiplet, $s$ is the 
flavor charge associated with the hypermultiplet, and $a$ and 
$a_D$ are functions of the modulus $u=\langle\Tr \,\Phi^2\rangle$.  
The \none chiral superfield $\Phi$ is the \ntwo
partner of the photon, and its $s$-charge is zero. 
The hypermultiplet is described by two superfields 
$\Phi_s$ for which $s=\pm 1$. 

This UV finite theory represents \nfour  SU(2) SYM at $m_h=0$. 
When $m_h$ is finite the theory flows in the infrared
to \ntwo SU(2) SYM, where $m_h$ plays the role of
a UV cut off while the infrared scale $\Lambda$ is
\begin{equation}
\Lambda^4=4\,m_h^4 \exp\left(2\pi \,i\,\tau_0\right)\,,\qquad
\tau_0=\frac{4\pi i}{g_0^2}+\frac{\theta_0}{2\pi}
\,,
\end{equation}
where the normalization is chosen for convenience.

\subsection{\nfour SYM}
\label{sec:3A}

Defining \nfour SU(2) SYM  as the limit $m_h\to 0$ implies
that vevs of the hypermultiplet fields $\Phi_s$ must vanish. 
On the Coulomb branch of the \nfour theory it is always possible 
to pass to such an orientation by global 
SU(4) rotations in the space of scalar fields. 
There is no running of the coupling and the vevs are given by their
classical values $a=\sqrt{2u}\,$, $a_D=\tau_0 a\,$. The central 
charge (\ref{zs2}) then takes the form (\ref{Zsc}),
\begin{equation}
{\cal Z}_{\{n_E,n_M\}}=\sqrt{2}\,a \left(\, n_E + n_M \,\tau_{0}\,\right)\,.
\end{equation}

The general expression (\ref{scalp1}) for the long range potential  
in this case results in
\begin{equation}
V(r)=-\,\frac{1-\cos \omega}{r\,{\rm Im}\,\tau_{0}}
\left|\, n_E^{(1)} + n_M^{(1)} \,\tau_{0}\,\right| 
\left|\, n_E^{(2)} + n_M^{(2)}
\,\tau_{0}\,\right|
\label{n4p}
\end{equation}
for the interaction between two BPS particles
with generic electric and magnetic charges
$\{n^{(i)}_E,\,n^{(i)}_M\}$.
This expression is valid for any nonzero $a\,$, while 
the point $a=0$ where all masses vanish should be excluded for the 
validity of our nonrelativistic approximation. 
We observe that the potential (\ref{n4p}) vanishes if the
central charges are aligned, i.e.
$\omega=0$, while it is attractive whenever they are misaligned.

In order to study the spectrum, let us start from 
the interaction of the  particles $\pm\{1,0\}$ and
$\pm\{0,1\}$. These are the lightest particles with nonvanishing electric and
magnetic charges which guarantees their stability, and thus
we will take these as ``primary'' constituents.  The potential (\ref{n4p})
leads to attraction in the channels $\pm\{1,1\}$,
$\pm\{1,-1\}$ but vanishes in the $\pm\{2,0\}$, $\pm\{0,2\}$ channels.
Therefore the potential leads to bound states 
with quantum numbers $\pm\{1,1\}$ and $\pm\{1,1\}$.
Although we are unable to use the nonrelativistic 
approximation to calculate the binding energy
--  it is of the same order as  the reduced mass -- we know that the ground
states in the $\pm\{1,1\}$, $\pm\{1,-1\}$ channels are indeed
BPS saturated from the corresponding dyon solutions at weak coupling. 

The vanishing of the $1/r$ potential in the $\pm\{2,0\}$ and 
$\pm\{0,2\}$ channels raises the question of whether
localized threshold states in these channels could appear due to 
subleading corrections. However, there are compelling arguments
against such a scenario. Firstly, attractive long range 
potentials, i.e. of ${\cal O}(1/r^2)$, are forbidden by the 
constraint that the mass cannot pass below the
BPS bound, as such potentials would always lead to a nonzero binding
energy. Secondly, bound states formed via short range interactions
are inconsistent with the moduli space formulation of 
multi-monopole dynamics, valid at weak coupling, where in 
these channels the relative separation is an exact modulus, and 
bound states are not expected \cite{sen}.

Knowledge of the long range potential between the two primary
states allows us to deduce the full spectrum of stable bound states
in the following way. Given that interactions between the primary
states are either attractive, i.e. between $\pm\{1,0\}$ and
$\pm\{0,1\}$, or vanish, we can ask whether the
constituents of a given composite can be arranged into two or 
more subgroups which do not interact. If the answer is positive, then
the two subgroups can dissociate without any cost in energy and
consequently a bound state is not formed by the $1/r$ potential.
On the other hand if the answer is negative, the constituents cannot
be separated and we have a bound state.

For a generic configuration with charges $\{n_E,n_M\}$, it follows
that we can arrange the constituents into $k$ non-interacting
subgroups only if the charges have a common divisor $k$, i.e.
they have the form $\{n_E,n_M\}=k\,\{n'_E,n'_M\}$ for some integer
$k$. Under the assumption that the bound states we find are indeed 
BPS, which we can verify in the semiclassical region 
for a restricted subset of charges, we then deduce that the stable 
BPS states form the set $\{n_E,n_M\}$ where $n_E$ and $n_M$ are co-prime.
Thus, we have obtained a spectrum of BPS states 
which is in agreement with known semiclassical 
results \cite{sen}, and indeed with the 
predictions of SL(2,{\bf Z}) duality \cite{mo,sen,f2}. We will
show that similar arguments can be applied to deduce the spectrum
in the \ntwo theory.

\subsection{Bound states and dissociation at strong coupling in \ntwo SU(2)}
\label{sec:3B}

In \nfour SU(2) SYM, since the gauge coupling is a marginal parameter,
we can choose to stay in the weak coupling regime. However, as 
we have emphasized, our approach is not limited to weak
coupling as we can make use of the Seiberg-Witten
solution for \ntwo SU(2) SYM to determine the long range interactions 
even when the microscopic gauge system is strongly coupled. 

Recall firstly that in the strong coupling region the BPS mass formula
cannot be specified globally due to the presence of cuts in the
relations $a(u)$ and $a_D(u)$ involving the global moduli space 
coordinate $u\,$. This implies that some states, while unique, require
a different specification of charges on each side of the 
cut (see e.g. \cite{fb1} for a careful analysis of this
issue). We will avoid this technical complication by always
restricting our attention to the upper half plane in the moduli 
space, ${\rm Im}(u)>0$, as para\-metrized by $u\,$, thus
avoiding the cuts which can be chosen to lie on the real axis.

\begin{figure}[ht]
\epsfxsize=6cm
\centerline{%
   \epsfbox{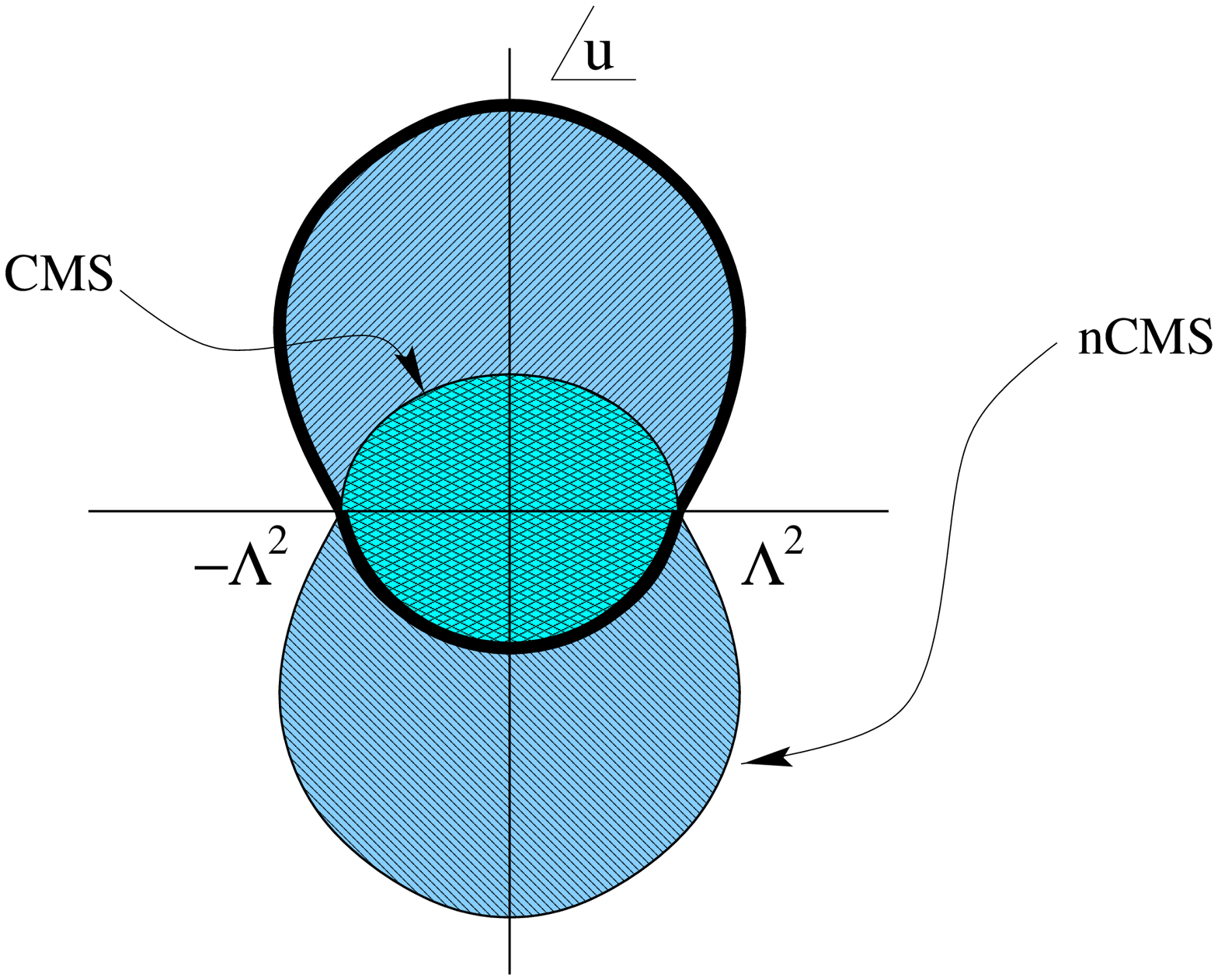}%
         }
 \vspace*{0.2in}
 \newcaption{The BPS curve of marginal stability (CMS) is shown in the
$u$-plane for SU(2) SYM, along with the non-BPS CMS (nCMS) where the
potential vanishes between certain states which both possess magnetic
charge. As an example, the bold contour bounds the exterior
stability domain for the non-BPS state with charges $\{1,2\}$.
\label{fi:Wboson}}
\end{figure}

To begin with we consider the interaction between two BPS particles
with generic electric and magnetic charges
$\{n^{(i)}_E,\,n^{(i)}_M\}$. The central charges of these states are 
\be
  \label{mas}
{\cal Z}_{i}=\sqrt{2}\,\left[n^{(i)}_E a(u) + n^{(i)}_M a_D (u)\right]\,.
\ee
The angle of disalignment $\omega$ is given by
 \begin{equation}
\label{phase}
{\rm e}^{i\omega}
=\frac{n^{(1)}_E+n^{(1)}_M\,
(a_D^*/a^*)}{\left|n^{(1)}_E+n^{(1)}_M\,(a_D/a)\right|}\;
\frac{n^{(2)}_E+n^{(2)}_M\, (a_D/a)}{
 \left|n^{(2)}_E+n^{(2)}_M\,
(a_D/a)\right| }\,.
\end{equation}
The long range interaction vanishes at $\omega=0$, 
i.e. when the central charges ${\cal Z}_{1}$ and ${\cal Z}_{2}$ 
are aligned,
\begin{equation}
{\rm Im}\left(\bar{\cal Z}_{1} {\cal Z}_{2}\right)=0\,,\qquad
{\rm Re}\left( \bar{\cal Z}_{1} {\cal Z}_{2}\right)>0\,.
\label{Zcon}
\end{equation}
These conditions are satisfied   all
over the moduli space when $\{n^{(1)}_E,\,n^{(1)}_M\}$ and
$\{n^{(2)}_E,\,n^{(2)}_M\}$ are parallel. This means that when 
the total electric ($n_E$) and magnetic ($n_M$) charges 
are {\em not} co-prime the situation is the same as it was in \nfour: 
the state dissociates into noninteracting BPS particles.
However, if $\{n^{(1)}_E,\,n^{(1)}_M\}$ and
$\{n^{(2)}_E,\,n^{(2)}_M\}$ are not parallel the phase $\omega$ becomes moduli
dependent. From the first condition in (\ref{Zcon}) it follows that the
modulus $u$ is on the CMS curve  defined by
\begin{equation}
{\rm Im} \left( a^*(u)\, a_D(u)\,\right)=0\,,
\label{cms}
\end{equation}
which is sketched in Fig.~\ref{fi:Wboson}.

Indeed, for $|\omega|\ll1$, we have from Eq.~(\ref{phase}) that
\begin{equation}
\omega={\rm Im}(a^*\,a_D) \,\frac{n_E^{(1)}n^{(2)}_M-n_M^{(1)}n^{(2)}_E}
{\left|\,n^{(1)}_E\,a+n^{(1)}_M\,a_D\right|  \left|\,n^{(2)}_E\,a +n^{(2)}_M\,
a_D\right|} +{\cal O}(\omega^2)\,.
\label{ome}
\end{equation}
This expansion is valid only when the second condition 
in Eq.~(\ref{Zcon}) is fulfilled, i.e. for
\begin{equation}
\left(n^{(1)}_E+n^{(1)}_M\, {\rm Re}(a_D/a)\right)
 \left(n^{(2)}_E+n^{(2)}_M\, {\rm
Re}(a_D/a)\right)>0\,,
\label{opi}
\end{equation}
otherwise, i.e. 
when ${\rm Re}\left( \bar{\cal Z}_{(1)} {\cal Z}_{(2)}\right)<0$, 
the angle  $\omega=\pi$ instead of zero on the curve (\ref{cms}). 
\noindent
Substituting the expression (\ref{ome}) for $\omega$ 
into Eq.~(\ref{Vcms}) we find the near-CMS potential
\begin{equation}
  \label{neCMS}
V_{\rm CMS}=-\frac{{\rm Im} (a^*\,a_D)}{r}\, 
\frac{\left(n^{(1)}_E n^{(2)}_M-n^{(1)}_M n^{(2)}_E\right)^2}
{\left|\,n^{(1)}_E\,a+n^{(1)}_M\,a_D\right|
\left|\,n^{(2)}_E\,a+n^{(2)}_M\,a_D\right|}
+{\cal O}(\omega^2)\,.
\end{equation}
The sign of the potential is determined by the
sign of ${\rm Im} (a^*a_D)$, and thus the problem of deducing the
stability domain for the relevant composite BPS states amounts 
to knowing the map between $a_D/a$ and $u$. In this example, the 
sign is easily seen to be positive in the exterior region by using the
semiclassical limit where $a_D=\tau a$ at large $u$. However, 
in more general situations it may prove useful to note that 
this problem can also be formulated as that of finding the
fundamental domain of ${\cal T}=a_D/a$. Recall that ${\cal T}$
transforms in the same manner as the coupling $\ta$ under the 
low energy monodromy group. This problem was solved for gauge group
SU(2) in \cite{afs} (see also \cite{matone}) with the conclusion 
that part of the fundamental domain of ${\cal T}$ 
(corresponding to the interior of the CMS (\ref{cms})) lies 
below the real axis. Thus we recover the picture
above and find that bound states are formed in the exterior 
region where the Coulombic coupling is attractive. 
However, upon crossing the CMS the sign flips and the
potential becomes repulsive in all channels satisfying the 
condition (\ref{opi}).

Let us follow construction of the spectrum in more detail.
In our discussion of \nfour SYM in the previous subsection, we chose
as primary states the W boson $\{1,0\}$ and the monopole $\{0,1\}$
as they are the lightest states carrying electric and magnetic charge
in the weak coupling region. It is clear that we can follow the same
procedure here provided we restrict our attention to the regime where
these states are the lightest in the relevant sectors. However,
we know that discontinuities in the spectrum will arise at strong
coupling and we need to consider again the appropriate choice of
primary particles in this region. One verifies that
on the CMS the ratio $a_D/a$ is real and, in the upper half
$u$-plane, it varies monotonically from $(-1)$ at $u=-\Lambda^2$ 
(the dyon $\{1,1\}$ is massless at this point) to 0 at 
$u=\Lambda^2$ (with a massless monopole $\{0,1\}$). It is clear then 
that the states $\pm\{0,1\}$ and $\pm\{1,1\}$ are stable BPS particles
over all of the moduli space. Thus in passing to strong coupling
the W boson is replaced by the $\{1,1\}$ dyon as the lightest
state carrying electric charge, which should then be chosen as a primary
particle. 

The W boson must then be understood as a composite configuration
and indeed we can consider the two particle system 
of $\{0,-1\}$ and $\{1,1\}$ which has the appropriate 
quantum numbers to form $W^+$. The condition (\ref{opi}) is satisfied in 
this case, so we arrive at the scenario described above: the $W$ boson
exists as a bound state only in the exterior  region of the CMS, 
its wavefunction swells on approach to the
CMS, the system becomes delocalized, and the bound state does 
not exist in the interior region. 
Note that this picture is not in contradiction
with the point-like nature of the
$W$ boson in the fundamental theory. At large moduli, 
$|u|\gg |\Lambda|^2$, we have a point-like $W$ boson with a 
small admixture of the dyon pair. As we move close to the CMS
this pair becomes dominant, and the picture of the $W$ boson as a
bound state becomes appropriate.

We can build up the spectrum of BPS states following the procedure
outlined for \nfour SYM. The new ingredient here is that we take
$\{0,1\}$ and $\{1,1\}$ as our primary constituents (in the upper 
half-plane). Moreover, the potential between these states may be
either attractive or repulsive depending on the choice of moduli.
Thus, for a given composite configuration, we can ask whether its
possible to arrange the primary constituents into subgroups 
in such a way that the energy is minimized when, for example, two of
these subgroups are at infinite separation. If so, then no bound state
is formed, while if the answer is negative, then we are guaranteed 
that a bound state is formed with those charges
at that point in the moduli space.

Lets consider first the possible composite configurations with
magnetic charge $\pm 1$. Following the above procedure, making use of the
potential (\ref{neCMS}), we obtain all the dyons $\pm\{k,1\}$ 
with $k=2,3,\ldots\,$ and $k=-1,-2,\dots\,$ as bound states in the exterior 
region to the CMS, while we find no bound states in the interior
region. The fact that these states are BPS is 
easily verified semiclassically from the moduli
space formulation, where they arise via quantization of the electric
charge, which is conjugate to one of the monopole moduli.
Together with the $W^\pm$ bosons these BPS particles form the 
well-known stable BPS spectrum \cite{sw,fb1} in the exterior 
region of the CMS.

\subsection{Non-BPS bound states in \ntwo SYM}
\label{sec:3C}

We now proceed further and consider composite configurations with
magnetic charge two. It is convenient to start with the simplest
example corresponding to a system with charges $\{1,2\}$.
We take states $\{0,1\}$ and $\{1,1\}$ as primary in the upper half plane.
The condition (\ref{opi}) is  not satisfied in this case, i.e. 
${\rm Re}\left( \bar{\cal Z}^{(1)} {\cal Z}^{(2)}\right)<0$ on 
the curve (\ref{cms}).
This means that $\omega=\pi$ and we cannot use the potential
(\ref{neCMS}), as there is no CMS for this state according to our
definition above. However, we can still study the full 
Coulombic potential,
\begin{equation}
 V_{\{1,2\}}(r) = \frac{1}{r\,{\rm Im}\,\tau}\;{\rm Re}\left\{\tau
 \left(1+\bar\tau\right)
 \left(1-\frac{a_D^*\,(a+a_D)}{|a_D|\,|a+a_D|}\right)
  \right\}, 
\label{21p}
\end{equation}
where
\begin{equation}
\tau(u)=\frac{\partial a_D}{\partial a}=\frac{4\pi i}{g^2(u)}+
\frac{\theta(u)}{2\pi}\,.
\end{equation}
In the weak coupling region where $|u|\gg |\Lambda|^2$, we have
$a_D=\tau a$ and this potential matches the \nfour potential
given in Eq.~(\ref{n4p}) with the substitution of $\tau_{0}$ by
$\tau$. The potential is attractive and the bound
state we deal with is a descendent of the \nfour BPS state with
magnetic charge two.

At first sight the appearance of states with magnetic charge 2 in \ntwo
SYM looks puzzling, as it is well known that such states do not exist
as BPS states in this theory,  i.e. one can argue following Bilal
and Ferrari \cite{fb1,bf2} that were such a state to exist it would
become massless on the CMS curve when $a_D/a=-1/2$ which contradicts
the Seiberg-Witten solution. The resolution is simple: 
the bound state we have found is {\em not} an \ntwo BPS state. 
At large $u$ this state can be viewed as a BPS state in
\nfour$\!$. However, the part of \nfour SUSY preserved by this state 
corresponds to generators carrying a nonzero value of $s$, i.e.   
involving fields from the \ntwo adjoint hypermultiplet. 
Thus, the supermultiplet we are dealing with here is short in terms
of \nfour supersymmetry but it is a ``long''  representation of 
\ntwo supersymmetry.  In the classical approximation, valid for 
large $u$, the bosonic configuration satisfies first order differential
equations and the soliton mass is given by the central charge. 
However, quantum corrections lift the
mass $M_{\{1,2\}}$ above the BPS bound $|{\cal
Z}_{\{1,2\}}|$. Nonetheless, the presence of an attractive potential
means that this mass is still less than the sum of the constituent masses,
\begin{equation}
\left|{\cal Z}_{\{1,2\}}\right| < M^{\rm non-BPS}_{\{1,2\}} <
\left|{\cal Z}_{\{0,1\}}\right|+\left|{\cal Z}_{\{1,1\}}\right|\,.
\label{massbnd}
\end{equation}

The potential (\ref{21p}) is of course also valid at strong coupling
and we can enquire as to the fate of these non-BPS states in this region.
In fact we find that in the upper-half plane 
the potential vanishes on a new curve, somewhat 
outside the CMS, defined by,
\be
 {\rm Re}\left\{\tau
 \left(1+\bar\tau\right)
 \left(1-\frac{a_D^*\,(a+a_D)}{|a_D|\,|a+a_D|}\right)
  \right\} = 0, 
\ee
which we will call a {\it non-BPS curve of marginal stability}
(nCMS) to emphasize the difference with the CMS curve. This
curve is plotted in Fig.~\ref{fi:Wboson}. Although we
cannot calculate the mass of the composite state on this curve there
are compelling arguments that the nCMS curve corresponds to the point
where the mass of the composite reaches the
threshold where the second inequality in (\ref{massbnd}) becomes
an equality, i.e. when
\be
 M^{\rm non-BPS}_{\{1,2\}}({\rm nCMS}) =
\left|{\cal Z}_{\{0,1\}}\right|+\left|{\cal Z}_{\{1,1\}}\right|\,.
\label{bndncms}
\ee
As evidence for this conclusion, note that if it is not true,
then the state would either disappear from the spectrum while stable
(i.e. $M^{\rm non-BPS}_{\{1,2\}} < |{\cal Z}_{\{0,1\}}|+|{\cal
Z}_{\{1,1\}}|$), or we could move to regions of the moduli space where
it was genuinely unstable to decay 
(i.e. $M^{\rm non-BPS}_{\{1,2\}} > |{\cal Z}_{\{0,1\}}|+|{\cal
Z}_{\{1,1\}}|$). To avoid both scenarios, we require that the nCMS
corresponds to the threshold condition (\ref{bndncms}).
Further progress beyond the nCMS then leads to a repulsive
potential and the non-BPS state consistently disappears from the spectrum.

We should emphasize that the position of the nCMS curve is not
corrected by higher derivative terms in the effective action.
This follows from its definition as the point where the 
$1/r$ potential between two BPS constituents vanishes, while 
the long range potential depends only on the two-derivative 
sector of the low energy effective theory. One consequence of this
is that the threshold condition (\ref{bndncms}) also receives no
corrections. Consequently, although we do not know the
mass of the non-BPS bound state for generic moduli, we know its value
precisely on the nCMS curve as it is given by the sum of the masses
of its two BPS constituents. Moreover, near this 
curve the constraint (\ref{const}) can always be satisfied and 
thus the nonrelativistic approximation is valid.

We can also deduce quite generally that the nCMS curve for a given
state must lie outside the CMS curve for the corresponding BPS state 
related to it by conjugation of the charges of one of its constituents. In 
particular, for the $\{1,2\}$ state we are considering here with
constituents $\{1,1\}$ and $\{0,1\}$ such a conjugate state is the 
W$^+$ boson $\{1,0\}$ with constituents $\{1,1\}$ and $\{0,-1\}$,
where we have conjugated the charge of the monopole. Denoting the
respective potentials $V_{\{1,2\}}$ and $V_{\{1,0\}}$, one verifies
that
\be
 V_{\{1,2\}} - V_{\{1,0\}} = \frac{2}{r\,{\rm
Im}\ta}\left(|\ta|^2-{\rm Re}\ta\right)\, > 0\,,
\ee
as follows from the positivity of Im$\ta$. This
result is not unexpected and is consistent with the fact that the
mass of the non-BPS composite lies above its BPS bound.

Thus far, we have restricted our discussion to the upper-half
$u$-plane, and have deduced that the state $\{1,2\}$ dissociates on 
an nCMS curve, as shown in Fig.~\ref{fi:Wboson}. The state 
$\{1,2\}$ also exists in the lower-half $u$-plane, for large $|u|$, and 
we can again ask what happens as we approach the strong coupling
domain. In the lower half-plane the only ``localized'' constituents
are $\pm\{0,1\}$ and $\pm\{-1,1\}$, which are necessarily 
the lightest in the strong coupling region. In terms of these constituents,
the state $\{1,2\}$ is actually a four-particle configuration comprised as 
follows:
\be
 \{1,2\} \Longleftrightarrow \{1,-1\} + 3\,\{0,1\}\,. 
\ee
This configuration can in principle dissociate in a more complicated
manner than the two-particle systems we have considered thus far.
However, the problem is still tractable following the procedure
discussed above. Noting that the monopoles do not interact, 
while the potential for $\{1,-1\}$ and $\{0,1\}$ is attractive 
outside, and repulsive inside, the CMS in the lower-half plane,
we find that the system can be separated into subgroups which minimize
their energy at infinite separation only in the interior
region to the CMS in the lower half plane. In fact, since the
monopoles do not interact, we can understand this result most easily
by noting that the wavefunction will take a simple product form
with the nontrivial structure depending only on the interaction
between $\{1,-1\}$ and $\{0,1\}$. Consequently, we 
deduce that the composite $\{1,2\}$ is present 
outside the CMS in the lower-half plane.

At first sight this seems puzzling as we have argued that the
composite $\{1,2\}$ is non-BPS and thus forms a long
multiplet. Specifically, the scenario we have found here 
corresponds to the case where both inequalities in (\ref{massbnd}) 
are saturated at the same point, implying that the CMS and the nCMS coincide.
Thus, precisely on the CMS we should only find BPS multiplets.
It follows that the transition to the CMS must lead to some degeneracy
as the state $\{1,2\}$ lies in a long multiplet arbitrarily close to
this point. In fact, this apparent problem is 
resolved by noting that the composite delocalizes precisely 
on the CMS itself, where it is replaced by the four-particle
configuration of BPS constituents, and we see that this indeed
provides the required degeneracy to account for the presence of
a long \ntwo multiplet.

Thus we have found a stability domain for the state $\{1,2\}$ 
which is given by the exterior region to 
the bold curve shown in Fig.~\ref{fi:Wboson}. 
The appearance of non-BPS states of this type, descending from higher charge
states in \nfour SYM, has been conjectured before in the literature
from the string web construction \cite{B2}. Moreover, 
the existence criteria for these non-BPS states, 
and also their stability domains, as deduced here
agrees well with earlier arguments of Bergman \cite{B2} using 
the string junction formalism.

Another state which we expect to be present at weak coupling
is related to $\{1,2\}$ by CP conjugation, namely the
state $\{-1,2\}$. In the lower half-plane there
are only two primary constituents for this composite
configuration, $\{-1,1\}+\{0,1\}$, and one
finds a stability domain bounded by the nCMS curve. In contrast,
in the upper half-plane there are four primary constituents, 
$\{-1,-1\}+3\{0,1\}$, and analysis of the possible two-body interactions
indicates that the composite is stable outside the CMS. Thus the
stability domain for the state $\{-1,2\}$ is the
reflection in the real $u$-axis of the domain for the state $\{1,2\}$.
This makes manifest a symmetry of the theory under CP conjugation
of the state and a Z$_2$ reflection in the moduli space
$u\rightarrow -u$ \cite{bf3}. In other words, we can now
see that the structure of CMS and nCMS curves 
in Fig.~\ref{fi:Wboson} is consistent with the
underlying Z$_2$ symmetry $u\rightarrow -u$ of the theory, 
due to the fact that it acts on the basepoint used in specifying 
the monodromies for each singularity and hence affects the charges. 
Consequently, to manifest this symmetry we need to consider 
both composite states $\{1,2\}$ and $\{-1,2\}$ that are 
related by just such a transformation of the basepoint. 

These examples provide sufficient information on the interaction
channels for us to deduce the presence, and stability domains,
for higher charge composite particles. In particular, we 
learn that all $\{n_E,n_M\}$ states with co-prime $n_E$ and 
$n_M$ -- descendants of \nfour -- represent 
states stable in the semiclassical region. These are, of course, 
{\em not} \ntwo BPS states: they lie in long \ntwo multiplets and 
their masses are larger than the values specified by the 
central charge in their charge sector. These states are removed from
the spectrum in the strong coupling regime as we cross the
appropriate nCMS or CMS curve. In regions of moduli space
where the latter case applies, the dissociation must respect the
degeneracies implicit in passing from long to short multiplets.

For the case of pure \ntwo SU(2) SYM that we have considered in this
section, the problem of constructing the spectrum of bound states
was simplified somewhat by the fact that there are only two 
primary states -- i.e. only two conserved charges, electric and
magnetic -- and thus deducing the existence of composite states could 
always be reduced to the problem of studying two-body interactions 
for which our long range potential is applicable. However, its clear 
that in more general scenarios, say for SU($N$) with $N\geq 3$, 
there will be more than two primary states, and thus more complex 
multi-body dissociations are possible for 
configurations carrying multiple charges. 
Nonetheless, one expects that at least for systems of a generic type 
the procedure outlined in this section will be sufficient to deduce the
presence or otherwise of composite configurations.

\section{Quark-Monopole Bound States in SU(2)}
\label{sec:4}

For our second example we turn to \ntwo SYM with gauge group SU(2)
with a single massive hypermultiplet in the fundamental
representation. This theory is the simplest example which exhibits
an Argyres-Douglas point \cite{ad,apsw} (see also \cite{gvy})
where two of the singularities in
the moduli space collide for a specific choice of the parameters.  
This phenomenon has interesting consequences for the CMS structure
of the theory.

For this system, the central charges take the form 
\begin{equation}
{\cal Z}_{\{n_E,n_M,s\}}=\sqrt{2}
\left[\,n_{E}\, a(u,m_f) + n_{M}\, a_{D}(u,m_f)\right] + s\, m_f\,, 
\label{zsf}
\end{equation}
where $m_f$ is the fundamental hypermultiplet mass.  
The quantum numbers of the fundamental fields, which we 
will refer to as quarks (see e.g. \cite{sw,f,kk} for a discussion of the
definition of the quark charges), are $n_E=\pm 1/2$, $n_M=0$, and $s=\pm 1$. 
For orientation on the quark masses it is convenient to start with  
large $m_f$, i.e.  $m_f\gg \Lambda_1$,
where $\Lambda_1$ is the scale parameter of the theory, and we have
used a chiral rotation to make $m_f$ real and positive. 
Then we find that the quark states ${\bf n}=\pm\{1/2,0,1\}$ with mass
$|\,m_f+a/\sqrt{2}\,|$ are heavier than  the quark states 
${\bf n}=\pm\{-1/2,0,1\}$ whose mass is $|\,m_f-a/\sqrt{2}\,|$. 
Moreover, for $u\approx m_f^2$ the lighter states ${\bf
n}=\pm\{-1/2,0,1\}$ actually become massless. This is the quark
singularity of the low energy effective description.

\subsection{Semiclassical regime and the CMS}

Following our discussion in the previous section, we will start by
considering the weak coupling regime, $|u|\gg |\La^2_1|$, where
the classical approximation is valid. The features of interest are
still visible in this region provided we also 
choose $m_f\gg \Lambda_1$ so that the quark singularity lies at 
weak coupling.

To construct the long range potential, we need to determine the
lightest BPS states carrying electric, magnetic, and quark charge.
For $m_f\gg \Lambda_1$ it it is easy to verify that the lightest such
states are the monopoles ${\bf n}^{(1)}=\{0,1,0\}$ and the ``light'' quarks 
${\bf n}^{(2)}=\{-s/2,0,s\}$ where
$s=\pm 1\,$. 
The central charges for these states take the form
\begin{equation}
{\cal Z}_{1}=\sqrt{2}\,a_D\,, \qquad {\cal Z}_{2}=s\left(m_f -
\frac{a}{\sqrt{2}}\right)
\end{equation}
and the long range interaction between them is
\begin{equation}
V_s=-\frac{1}{2\,r\,{\rm Im}\tau}\,{\rm Re}\left\{\tau\,\frac{ a_D^*
(a-m_f\sqrt{2})}{|a_D||a-m_f\sqrt{2}|}+s\,\tau\right\}\,.
\label{Vs}
\end{equation}
At leading order in the weak coupling regime, where $a=\sqrt{2u}$ and 
$a_D=\ta a$, the potential is independent of $s$
and takes the simple form,
\be
 V_s^{\rm cl} =  -\,\frac{1-{\rm
Re}(m_f/\sqrt{u})}{2\,r\left|\,1-\left(m_f/\sqrt{u}\right)\right|}
\,.
\label{Vqmsc}
\ee
We deduce that the potential is attractive for Re$(m_f/\sqrt{u})<1$
and thus there are dyonic bound states with charges $\{-s/2,1,s\}$
for $s=\pm 1$ in this region.  This conclusion agrees precisely with 
an analysis of the Dirac equation in the monopole background 
\cite{henn}, which provides a useful check on our result. 

On closer inspection we find that this result is actually stronger
than that predicted by an analysis of the CMS conditions (\ref{Zcon}).
Alignment of the central charges implies 
\begin{equation}
{\rm Re}\left(\frac{m_f}{\sqrt{u}}\right)=1\,,\qquad
s\,{\rm Im}\left(\frac{m_f}{\sqrt{u}}\right)\ge 0\,. \label{qmsccms}
\end{equation}
Thus, if we restrict our attention here to the upper-half $u$-plane,
we see that the CMS constraint is only satisfied for $s=-1$. Hence the
transition in the potential (\ref{Vqmsc}) at Re$(m_f/\sqrt{u})=1$
only corresponds to crossing a CMS for the state $\{1/2,1,-1\}$. In
contrast the constituent central charges are anti-parallel for the 
state $\{-1/2,1,1\}$ at this point. We conclude that this second state
is actually a non-BPS bound state. In the classical approximation that
we are using the CMS and nCMS curves for these two states coincide,
and thus this result is consistent with \cite{henn}. However, once
quantum corrections are included, we would expect these curves to
split in accordance with corrections to the mass of the $\{-1/2,1,1\}$
state above its BPS bound.

This splitting can be verified from Eq.~(\ref{Vs}) by taking into
account the perturbative corrections due to the 
running of $\tau$, which provide the leading $s$ dependence 
of the potential at weak coupling. The perturbative running 
coupling $\tau$ is given by
\begin{equation}
\tau(u)=\frac{i}{\pi}\log\frac{u}{\Lambda_0^2}\,,\qquad
\Lambda_0=(m_f\Lambda_1^3)^{1/4}\,,
\end{equation}
where $\Lambda_0$ is 
the scale of the effective $N_f\!=\!0$ theory. Then we find
\be
 V_{s=1}- V_{s=-1} =\frac{1}{r}\frac{{\rm Arg}\,
u}{\log\left(|u|/\Lambda_0^2\right)}>0\,,
\ee
which indicates that the instability domain for the non-BPS bound state
$\{-1/2,1,1\}$ is wider than that for the BPS state $\{1/2,1,-1\}$.

\subsection{Strong coupling and the Argyres-Douglas point}

\begin{figure}[ht]
\centerline{%
   \epsfxsize=7cm\epsfbox{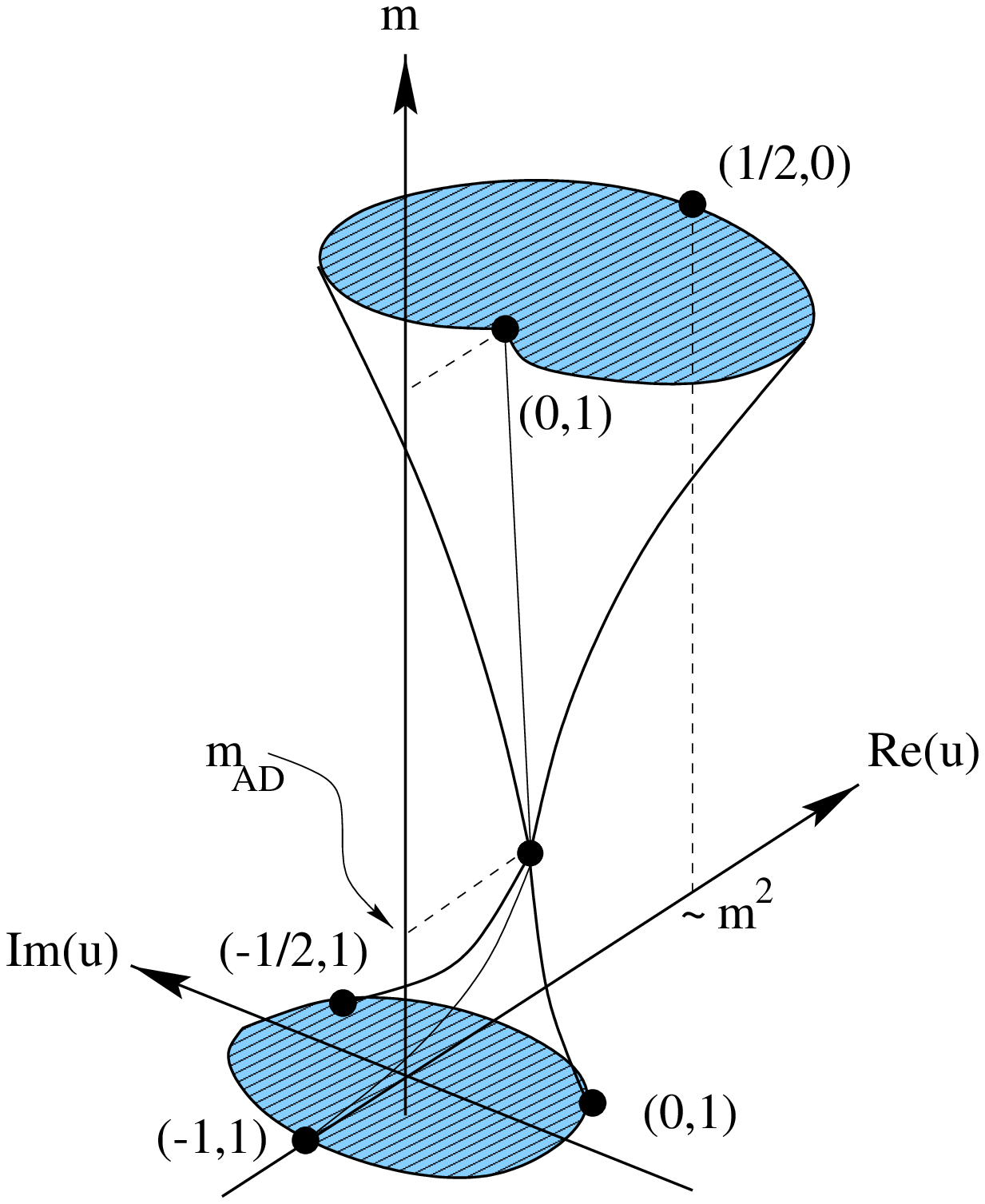}%
         }
 \vspace*{0.2in}
 \newcaption{The CMS curve is shown as a function of $m=m_f$  for the state
$\{1/2,1,-1\}$. For $m_f\gg m_{AD}$,  the CMS takes the form of an approximate
cardioid encompassing the monopole ($u \approx \La_0^2$) and 
quark singularities ($u\approx m_f^2$). As $m_f\rightarrow m_{AD}$ 
this region shrinks to a point and while for $m_f=0$ it becomes 
the massless CMS curve ${\rm Im}(a_D^* a)=0$. The singularities
are labeled with their charges.
\label{fig:zeroqm}}
\end{figure}

We now return to a more detailed analysis of the CMS structure for the
BPS state $\{1/2,1,-1\}$. The alignment
conditions (\ref{qmsccms}) have the general form,
\be
 {\rm Im}\left[a_D^* a \left(\frac{m_f\sqrt{2}}{a}-1\right)\right]=0\,,\qquad
s\,{\rm Re}\left[a_D^* a \left(\frac{m_f\sqrt{2}}{a}-1\right)\right]\ge 0\,.
\label{qmcms}
\ee
defining a CMS curve which by definition 
must pass through the quark and monopole singularities.
Using explicit expressions for the period integrals (see
e.g. \cite{amz,bf3}), we can go beyond the weak coupling regime and
construct the CMS also in the strong coupling region. We find that
it forms a closed curve which approximates a cardioid, and encompasses
the quark singularity at $u\approx m^2$ and the monopole singularity
at $u\approx \Lambda_0^2$.  This curve is
shown in Fig.~\ref{fig:zeroqm} as a section at fixed $m$ of the
full instability domain for the state $\{1/2,1,-1\}$ in the  
three dimensional space  $\left[\,{\rm Re}(u),\,{\rm
Im}(u),\,m_f\,\right]$. 

Thus far we have considered the regime where $m_f\gg \La_1$. It is 
clear that the picture remains qualitatively the same until we
approach the Argyres-Douglas point at $m=m_{AD}=(3/4)\Lambda_1$
where the monopole and quark singularities coalesce
and the CMS we observed above shrinks to a point.  
When $m_f$ is reduced below $m_{AD}$ the quark singularity undergoes
a monodromy as we unwind the cuts and is relabeled with the charges
$\{-1/2,1,1\}$ (not to be confused with the non-BPS state we found above).
The state $\{1/2,1,-1\}$ is still composite and on 
passing through the Argyres-Douglas point
we find that a new CMS opens up in the strong coupling region
encompassing the singularities of the monopole, which moves into the
lower-half plane, and the continuation of the quark, now 
labeled as a $\{-1/2,1,1\}$ dyon, which moves into the upper-half plane. 
The $\{1/2,1,-1\}$ state is excluded from the interior of this
region. As $m_f\rightarrow 0$ this CMS expands and precisely when 
$m_f=0$ it reaches an approximately circular form, given by
${\rm Im}(a_D^* a)=0$ (see Eq.~(\ref{qmcms})), and includes the
third dyon singularity which in the upper-half plane
has charges $\{-1,1,-1\}$ (NB: these charges correspond to a
specific arrangement of the cuts of $a(u)$ and $a_D(u)$ consistent
with \cite{bf2} for $m_f=0$). The three
singularities then form the well-known {\bf Z}$_3$ symmetric formation
of the massless theory \cite{sw}. Thus we can now represent the
full instability domain for the state  $\{1/2,1,-1\}$,
which we present in Fig.~\ref{fig:zeroqm},  including the
regime $m<m_{AD}$.

Thus, we have found that the
behavior of the potential is entirely consistent with earlier 
arguments in the literature regarding the BPS spectrum at 
weak coupling \cite{ssz,gh1,henn} and on flowing through the Argyres-Douglas
point \cite{bf2,bs,bf3}. 
Moreover, we have also deduced the presence of non-BPS dyonic
bound states with quark charge which arise in a similar manner to 
our discussion in the previous section.
Although we have considered a rather specific example, it seems 
clear that similar phenomena will arise in considering other states. 
In the next section, we generalize somewhat further and consider 
higher rank gauge groups.

\section{Dyons in SU(3) SYM}
\label{sec:5}

New phenomena are possible when the rank of the
gauge group is extended beyond 1. In particular, it will be sufficient
here to focus on the simplest nontrivial example -- an SU(3) gauge
group. For \ntwo SYM with gauge group SU(3) and one 
adjoint hypermultiplet of mass $m_h$, the central charges are
\begin{equation}
 {\cal Z}_{\{n^{a}_{E},n_{Ma},s\}} = \sqrt{2}\,\left[\,n^{a}_{E} a_a + n_{Ma}
a^a_{D}\,\right] +s\,m_h \,.
\end{equation}
The integral electric charges $n^a_{E}$ ($a=1,2$) are associated 
with the two simple roots $\bbox{\beta}_a$ of SU(3) while the 
magnetic charges $n_{Ma}$ are associated with the corresponding
dual vectors, namely the weights $\bbox{\om}^a=C^{ab}\bbox{\beta}_b$, 
where $C^{ab}$ is the inverse Cartan matrix which in our normalization
takes the form,
\begin{equation}
C_{ab}=\bbox{\beta}_a\cdot \bbox{\beta}_b=\left(
\begin{array}{c c}
1&-1/2\\
-1/2&1
\end{array}
\right)\,,\qquad
C^{ab}=\bbox{\om}^a\cdot \bbox{\om}^{b}=\frac{4}{3}\left(
\begin{array}{c c}
1&1/2\\
1/2&1
\end{array}
\right)\,.
\end{equation}
This theory is by construction UV finite and taking $m_h=0$ we obtain
\nfour SYM. A finite value for $m_h$ acts as a UV cutoff in the
\ntwo theory which flows in the infrared to \ntwo SYM.

\subsection{\nfour SYM}

Let us start with \nfour SYM, i.e. by taking $m_h=0$. 
In this case $a_a=\langle \bbox{\Phi}\cdot \bbox{\beta_a}\rangle$ 
and $a_D^a= \tau_0 \,C^{ab} a_b$. As discussed in Sec.~\ref{sec:2}, the
usual semiclassical treatment \cite{gno,weinberg}
makes use of magnetic charges $n^a_{M}$
with an upper index\,\footnote{Note that  the monopole 
$\{n_M^1=1,n_M^2=0\}$ aligned along $\bbox{\beta}_1$ has half integer charges
$n_{Ma}$ in the notation
of the dual lattice,  $\{n_{M1}=1,n_{M2}=-1/2\}$.},
$n^a_{M}\!=\!C^{ab}\,n_{Mb}$. The 
central charge  then takes the form (\ref{Zsc})
which we can also write as
\begin{equation}
 {\cal Z}_{\{n^{a}_{E},n_{M}^a\}} = \sqrt{2}\,Q^a\, a_a \,, \qquad
Q^a=n^{a}_{E} +\tau_0\, n_{M}^a\,,
\label{ctrl}
\end{equation}
where we have introduced the complex charges $Q^a$.

We consider the interaction between two fundamental dyons  
living in two different U(1) subgroups of the Cartan subalgebra,
namely dyons with quantum numbers $\{(n_E^1,0), (n_{M}^{1},0)\}$ 
and $\{(0, n_E^2), (0, n_{M}^{2})\}$ in the notation 
${\bf n}=\{(n_E^1,n_E^2),
(n_{M}^{1},n_{M}^{2})\}$ (note the lifted index for magnetic charges).
The central charges then take the form,
\be
 {\cal Z}_{1} = \sqrt{2}\,Q^{1}\, a_{1} \,,\qquad 
 {\cal Z}_{2} = \sqrt{2}\,Q^{2}\,a_{2}\,, 
 \label{Zwc}
\ee
and the potential, which follows from the general
expression (\ref{scalp1}), is in this case
\begin{eqnarray}
&& V(r) = \frac{C_{12}}{ r\, {\rm Im} \tau_0}\,|Q^{1}|\,|Q^{2}| \, {\rm Re}
\left\{\,
\frac{Q^1(Q^2)^*}{|Q^{1}||Q^{2}|} - 
\frac{(a_1)^* a_2}{|a_{1}||a_{2}|}\right\},
\label{Vwc}
\end{eqnarray}
where the the element of the Cartan matrix $C_{12}=-1/2$. 
Both terms in the curly brackets are phase factors,
\begin{equation}
\frac{Q^1(Q^2)^*}{|Q^{1}||Q^{2}|} ={\rm e}^{-i\omega_Q}\,,\qquad
\frac{(a_1)^* a_2}{|a_{1}||a_{2}|}={\rm e}^{i\omega_V}\,,
\end{equation}
where the first, $\omega_Q$, is the relative phase of the two charges, 
and the second, $\omega_V$, is the relative phase of the 
two vevs, $a_a=\langle\bbox{\Phi}\cdot \bbox{\beta_a}\rangle$. The 
sum of $\omega_Q$ and $\omega_V$ is the relative
phase of the central charges introduced in Eq.~(\ref{omd}), 
$\omega=\omega_Q+\omega_V$. In terms of these phases the potential 
is given by
\begin{equation}
V(r) =-\,\frac{1}{ r\, {\rm Im} \tau_0}\,|Q^{1}|\,|Q^{2}|
\,\sin\frac{\omega_V+\omega_Q}{2}\,\sin\frac{\omega_V-\omega_Q}{2}\,.
\label{3pot}
\end{equation}
If the complex charges $Q^{1,2}$ are aligned, i.e. $\omega_Q=0$, 
the potential (\ref{3pot}) is negative semi-definite over the moduli
space,  while it is positive semi-definite  if $Q^{1,2}$ are anti-aligned,
i.e $\omega_Q=\pi$.  

The most interesting situation arises when $Q^{1,2}$ are not aligned.
If, for example, we choose $n_M^1=n_M^2=1$ then at weak coupling the phase
$\omega_Q$ is small and proportional to the difference of 
the electric charges, 
\begin{equation}
\omega_Q=\frac{g_0^2}{4\pi}\,(n_E^1-n_E^2)\,.
\end{equation}
The potential is repulsive in the domain of the moduli space where
$|\omega_V|<|\omega_Q|$, implying that there are no bound states with
$|n_E^1-n_E^2|>4\pi|\omega_V|/g_0^2$. However, the potential
is attractive for $|n_E^1-n_E^2|<4\pi|\omega_V|/g_0^2$ and 
consequently leads to the formation of bound states in this region.

When $\omega_V=\pm \omega_Q$ the Coulombic potential vanishes. 
Specifically, the relation $\omega_V=- \omega_Q$ is 
equivalent to the vanishing of $\omega=\omega_Q+\omega_V$ which implies that  
this is the CMS where the central charges ${\cal Z}_{1,2}$ are aligned. 
However, the vanishing of the potential at
$\omega_V=\omega_Q$ cannot be interpreted in this way. The resolution
of this point in the current system is that there is a second CMS
curve related to the presence of a second central charge in the
\nfour SUSY algebra. This second charge, $\tilde{\cal Z}$, has the form
\begin{equation}
\tilde{\cal Z}_{\{n^{a}_{E},n_{M}^a\}} = \sqrt{2}\,Q^a\, a_a^*\,,\qquad
\left|\tilde{\cal Z}_{\{n^{a}_{E},n_{M}^a\}}\right|=
\left|{\cal Z}_{\{-n^{a}_{E},n_{M}^a\}}\right|\,,
 \label{Ztilde}
\end{equation}
where we have also noted that $|{\cal Z}|$ and $|\tilde{\cal Z}|$ are
related by electric charge conjugation. As is clear from comparison of
Eqs.~(\ref{ctrl}) and (\ref{Ztilde}) the distinction between ${\cal Z}$ and 
$\tilde{\cal Z}$ is significant only for gauge 
groups of rank greater than one. In the
present context we see that
the CMS for the $\tilde{\cal Z}$ central charge corresponds
to  $\tilde \omega=\omega_Q-\omega_V\to 0$, which explains the fact that
we found a vanishing potential for $\omega_V=\omega_Q$.

For the bound states we are discussing, $|{\cal Z}|\neq  |\tilde{\cal
Z}|$, and thus they preserve 1/4 of the \nfour SUSY. In recent years
a considerable amount of work has been undertaken on these states in the
semiclassical 
regime \cite{ly,bhlms,tong,blly,bly,bl,bly2,n2quant,rsvv,gkly}, and 
also in the context of 
string junctions \cite{bergman,bk}. 
In the semiclassical regime, a more detailed
picture of the dynamics has recently been obtained
\cite{ly,tong,bly,bly2,n2quant,gkly}
using the moduli space 
approximation \cite{mgeo}, 
while the $1/r$ potential for the SU(3) theory was
also discussed in this regime in Refs.~\cite{lwy2,fh2,blly,rsvv}. 
Our main point here is that a simple analysis of the 
Coulombic forces is sufficient
for finding the spectrum of BPS states and also the rich structure of 
CMS curves which arise for higher rank gauge groups.

\subsection{\ntwo SYM in the semiclassical range}

By giving a mass $m_h$ to the hypermultiplet fields $\Phi_s$, we can 
flow to \ntwo SYM and observe how the picture above is corrected
at scales well below $m_h$. The generic expression for the 
near-CMS potential, as given in (\ref{Vcms}), determines CMS curves at
\be
 {\rm Im}\left[(n_E^{(1)a}a_a^* + n_{Ma}^{(1)}a_{D}^{a*})
 (n_E^{(2)b}a_b + n_{Mb}^{(2)}a_{D}^{b})\right]=0.
\ee
The solution set of this equation is quite complex
and we will not pursue an in-depth 
analysis of the potential here, although the behavior 
is not unlike that of the one-flavor SU(2) case considered above. 
In particular, this theory possesses an Argyres-Douglas point
\cite{ad} of the same type \cite{apsw}.

To simplify the discussion, we will limit ourselves to the
semiclassical range of large moduli. In this region the long range 
potential is close to the one found above in the \nfour
case modulo small corrections due to the running of the gauge
couplings.  This means that in the weakly coupled range of the \ntwo
theory we necessarily have bound states for the same set
of charges $\{n_E^a,n_M^a\}$ as in \nfour SYM.  However, as we have
observed in the SU(2) examples, not all these states are BPS saturated
from the point of view of \ntwo SUSY. 

The example considered in detail above, involving the interaction of the states
$\{(n_E^1,0), (1,0)\}$ and $\{(0, n_E^2), (0,1)\}$, is
sufficient to understand how the splitting of the spectrum arises.
Indeed, as described above, there are two CMS curves for this system
in \nfour SYM: one, $\omega_V=-\omega_Q$, where the central charges ${\cal
Z}_{1,2}$ are aligned; and the other, $\omega_V=\omega_Q$, where the
additional \nfour central charges $\tilde{\cal Z}_{1,2}$ were aligned.
This second central charge does not lead to shortening of 
the \ntwo multiplets \cite{n2quant}, 
and so the bound states associated with the
second CMS in \nfour are not BPS states of \ntwo SYM. 
This implies that, say, for positive $\omega_V$ only those states 
in the interval $0<n_E^2-n_E^1< 4\pi \omega_V/g^2$ are BPS states of 
\ntwo lying in short multiplets. The states with the opposite sign 
of $n_E^{1,2}$ do exist as bound states in the semiclassical
region when $-4\pi \omega_V/g^2 < n_E^2-n_E^1 < 0$
\cite{n2quant,rsvv},  but they are in long multiplets of 
\ntwo and their masses are larger than the BPS bound $|{\cal Z}|$, 
where ${\cal Z} =  {\cal Z}_{1}+{\cal Z}_2$ is the total
central charge. This is consistent 
with the relation $|\tilde{\cal Z}|>|{\cal Z}|$
near the second CMS, discussed in \cite{n2quant}.

\section{Concluding Remarks}
\label{sec:6}

In this paper we have argued that the generic features of the
spectrum of composite BPS states in \nfour and \ntwo SYM may be deduced purely
from knowledge of the long range potential between primary constituents.
Moreover, we showed that this potential was calculable given knowledge
of the low energy effective theory describing the massless
sector. This result is valid for generic field theoretic moduli and
holds in particular in the strong coupling region due to the
nonrelativistic dynamics in the near CMS regime. This limit also
leads to the moduli space geometry becoming irrelevant near the CMS, 
where the Coulombic potential was found to have a remarkably
simple form, independent of the K\"ahler potential.

We dwelt on several illustrative examples in the context of
\ntwo SYM with gauge groups SU(2) and SU(3), and from this analysis
we can construct a rather complete picture of the BPS spectrum in 
\ntwo theories. Namely, we start from the weak coupling regime where
the Coulombic potential will predict a spectrum of bound states, 
the presence of many of which can be verified independently by 
semiclassical techniques. We can then follow these states on 
trajectories through the moduli space to strong coupling. If the 
states are BPS states in \ntwo then we must either encounter a CMS
for the state, or it must be massless somewhere in the moduli space.
These conditions are easily checked given knowledge of the
central charge. In the absence of either of these scenarios, 
we can conclude that the semiclassical bound state was non-BPS and 
will generically disappear from the spectrum at curves which we called
the nCMS  before  reaching the CMS. For special trajectories in the
moduli space, it may happen that the nCMS and the CMS coincide, but we
found no examples where this occurred for all possible trajectories,
implying that BPS and non-BPS states are indeed distinguished via
this procedure. For BPS states, this picture is
closely related to similar arguments by Bilal and
Ferrari \cite{fb1,bf2,bf3}. What we have added here is that some of
the working assumptions of \cite{fb1,bf2,bf3} can be deduced directly
from an analysis of the Coulombic potential.

In general, the non-BPS states predicted by the Coulombic potential in the
examples of the preceding sections have a larger instability domain
than the BPS states to which they are related by various conjugations
of the charges. It would be interesting to explore the presence of
these states in more detail. 
In this regard, we
note that higher charge non-BPS states in \ntwo have been predicted via 
the analysis of string webs \cite{bk,B2}. Moreover,
in these string web constructions, a version of the
``s-rule'' \cite{Wsj} is invoked to determine which  
states are BPS states in \ntwo$\!$. In our approach this is related
to the alignment condition, Re$(\bar{\cal Z}_1{\cal Z}_2)>0$. 
However, it would be interesting
to know if there is a constraint of this kind
which is applicable locally on the Coulomb branch, as is 
suggested by certain string junction constructions \cite{Wsj}\,\footnote{We
thank P. Argyres for helpful comments on these 
and related issues.}. 

Another issue concerns those states for which the Coulombic
potential behaves as ${\cal O}(\om^2)$ near the CMS. We plan to
discuss these systems elsewhere \cite{rv2}, but we note that examples of
states of this kind, accessible in the semiclassical region, have
stability domains on both sides of the CMS and a vanishing 
equilibrium separation between the constituents at the classical
level. These include states with purely electric or purely magnetic
charges, as mentioned in Sec.~\ref{sec:5}. When the Higgs vevs are
aligned these states have been argued to exist as threshold bound states 
in \nfour SYM \cite{gl,lwy,gibbons}, but
are not present as BPS states in \ntwo SYM \cite{n2quant,sy}.

\bigskip
\subsection*{\bf Acknowledgments}
\addcontentsline{toc}{section}{\numberline{}Acknowledgments}
We are grateful to Philip Argyres, Gregory Gabadadze, 
Mikhail Shifman, Mikhail Voloshin,  Mikhail Olshanetsky, and
Alexei Yung for helpful discussions, and we thank Philip Argyres and
K. Narayan for helpful comments on a preliminary draft of this paper. 
This work was supported by the DOE under grant number DE-FG02-94ER408.

\newpage

\end{document}